\documentclass[aip, jcp, citeautoscript,amsmath,amssymb,reprint]{revtex4-2}

\PassOptionsToPackage{colorlinks=true,allcolors=blue}{hyperref}

\usepackage{graphicx}
\usepackage{comment}
\usepackage{dcolumn}
\usepackage{bm}
\usepackage[version=4]{mhchem}

\usepackage[table,dvipsnames]{xcolor}
\usepackage{colortbl}
\usepackage{tikz}

\usepackage{chngcntr}
\usepackage{hyperref}
\usepackage{orcidlink}

\hypersetup{
  colorlinks=true,
  allcolors=blue
}
\AtBeginDocument{\nocite{achemso-control}}

\usepackage[utf8]{inputenc}
\usepackage[T1]{fontenc}

\usepackage{mathptmx} 
\usepackage{threeparttable}

\begin{document}

\title{
Coupling between thermochemical contributions of subvalence correlation and of higher-order post-CCSD(T) correlation effects --- a step toward `W5 theory' 
}

\author{Aditya Barman\orcidlink{0009-0003-3863-2564}} 
\affiliation{
  Department of Molecular Chemistry and Materials Science, Weizmann Institute of Science, 7610001 Re\d{h}ovot, Israel.
}
\author{Gregory H. Jones\orcidlink{0000-0003-3275-1661}}

\affiliation{Quantum Theory Project, Department of Chemistry,
University of Florida, Gainesville, FL 32611, USA}
\author{Kaila E. Weflen\orcidlink{0009-0003-2483-8692}} 
\affiliation{
  Department of Molecular Chemistry and Materials Science, Weizmann Institute of Science, 7610001 Re\d{h}ovot, Israel.
}
\author{Margarita Shepelenko\orcidlink{0000-0003-4707-1650}} 
\affiliation{
  Department of Molecular Chemistry and Materials Science, Weizmann Institute of Science, 7610001 Re\d{h}ovot, Israel.
}
\author{Jan M. L. Martin\orcidlink{0000-0002-0005-5074}} 
 \email[Corresponding author: ]{gershom@weizmann.ac.il}
\affiliation{
  Department of Molecular Chemistry and Materials Science, Weizmann Institute of Science, 7610001 Re\d{h}ovot, Israel.
}

\date{\today}

\begin{abstract}
We consider the thermochemical impact of post-CCSD(T) contributions to the total atomization energy (TAE, the sum of all bond energies) of first- and second-row molecules, and specifically their coupling with the subvalence correlation contribution. In particular, we find large contributions from (Q) when there are several neighboring second-row atoms. Otherwise, both higher-order triples $T_3$–(T) and connected quadruples (Q) are important in systems with strong static correlation. Reoptimization of the reference geometry for core-valence correlation increases the calculated TAE across the board, most pronouncedly so for second-row compounds with neighboring second-row atoms. 

We present a first proposal for a "W5 theory" protocol and compare computed TAEs for the W4-08 benchmark with prior reference values. For some key second-row species, the new values represent nontrivial revisions. Our predicted TAE$_0$ values (TAE at 0 K) agree well with the ATcT (active thermochemical tables) values, including for the very recent expansion of the ATcT network to boron, silicon, and sulfur compounds.
\end{abstract}

\maketitle

\maketitle
\makeatletter
\@removefromreset{table}{section}
\@removefromreset{table}{subsection}
\renewcommand{\thetable}{\Roman{table}}
\makeatother

\section{\label{sec:intro}Introduction }

Accurate thermochemistry is seeing a modest renaissance in recent years. This is thanks to the tandem efforts of, on the one hand, the ATcT (Active Thermochemical Tables\cite{ATcT,ATcTpaper1,ATcTreview,Ruscic2019}) team at Argonne National Laboratory, and, on the other hand, the development of high-accuracy computational protocol families such as HEAT by an international consortium centered around the late lamented John F. Stanton,\cite{HEAT,HEAT2,HEAT3,HEAT4} Weizmann-4 (W4) theory developed by our group,\cite{jmlm200,jmlm205,jmlm330} and the FPD (Feller-Peterson-Dixon) strategy.\cite{FPDreview1,FPDreview2} For a recent review, see Karton\cite{Karton2022}.

These latter techniques (and to a lesser extent, reduced-cost variants such as W4-F12\cite{jmlm269,jmlm285},  W3X-L,\cite{Chan2015} and Wn-P34\cite{Chan2021}) now offer kJ/mol accuracy on a semi-routine basis.

Of late, the ATcT project has been moving into the second row of the Periodic Table, and hence the ATcT team expressed a desire for ab initio TAEs (total atomization energies) of second-row species accurate to 1 kJ/mol (0.24 kcal/mol; 83.59 cm$^{-1}$) or better. The W4-17 benchmark\cite{jmlm273} contains a fair amount of second-row species --- but owing to computational cost and resources limitations, any post-CCSD(T) corrections were limited to the valence electrons, aside from W4.4 data\cite{jmlm205} for a handful of second-row diatomics (such as \ce{Cl2} and \ce{S2}).

HEAT\cite{HEAT,HEAT2,HEAT3,HEAT4} was developed with first-row systems in mind and makes no effort to separate valence from inner-shell correlation (a.k.a., subvalence correlation), while W4 and its predecessors\cite{jmlm127,jmlm148,jmlm151,jmlm173} were aiming at second-row systems from the start and hence, of necessity, separated valence and inner-shell correlation from the ground up. Still, otherwise HEAT and Weizmann-$n$ have pretty much converged to each other.\cite{HEAT3}

In Ref.\cite{jmlm281}, subvalence correlation was considered near the complete basis set limit at the CCSD(T) level using three different basis set families. Among other things, the authors of that study found that first-row and second-row molecules behave qualitatively differently. While in first-row molecules, core-valence correlation predominates, and is almost exclusively attractive, in second-row molecules with multiple adjacent second-row atoms, a repulsive core-core contribution may partly compensate for an attractive core-valence contribution. (Triple substitutions were found to be attractive throughout.)

Neither W4 nor HEAT, in their unmodified form, treat post-CCSD(T) contributions to the inner-shell correlation, although W4.3 and W4.4 do so. At the time (2004-2008) the two `competing' approaches (and for that matter FPD) were developed, inclusion of such contributions would have been computationally intractable for second-row species beyond diatomics. Fortunately, two decades of hardware evolution have removed this obstacle. Therefore, in the present paper, we seek to address the importance of post-CCSD(T) subvalence correlation in detail.

We will also address two subsidiary questions that arise when comparing HEAT and W4-type approaches. 

(a) Both use CCSD(T) optimized reference geometries in correlation consistent\cite{Dunning1989,Dunning2002} quadruple-zeta basis sets. However, while W4 freezes subvalence electrons in the CCSD(T)/cc-pV(Q+d)Z geometry optimization (again, a choice made in the interest of being able to treat second-row species with limited computational resources), the original HEAT protocol \emph{did} include subvalence correlation.\footnote{For reasons of computational expediency, the original HEAT papers\cite{HEAT,HEAT2,HEAT3} prescribed a cc-pVQZ basis set for the all-electron optimization; evidently, as first pointed out by Taylor,\cite{Taylor1992} this is less than ideal considering the cc-pVQZ basis set is of just \emph{minimal basis} quality in the subvalence orbitals. Hence we used core-valence optimized basis sets. Now while for first-row elements, the difference between core-optimized cc-pCVQZ\cite{WoonDunning1995_pCVnZ} and core-valence weighted cc-pwCVQZ\cite{pwCVnZ} is essentially one of semantics, the latter are definitely favored for second-row elements.} The HEAT team raised the question to what extent this would affect computed TAEs, all else being equal.

(b) The HEAT team has traditionally preferred UHF (unrestricted Hartree-Fock) references, while previous Weizmann-$n$ editions eschewed it in favor of ROHF (restricted open-shell Hartree-Fock), sidestepping various spin contamination artifacts. If any subvalence orbitals are `frozen' (constrained to be doubly occupied), ROCCSD(T)\cite{Wat93} contains an ambiguity discussed in the  Appendix to Ref.\cite{jmlm200}: whether to semicanonicalize the ROHF orbitals after integral transformation (and any dropped cores, the MOLPRO\cite{MOLPRO} choice) or prior to it (the path followed in CFOUR\cite{CFOUR} and most other coupled cluster codes).
We will depart from past practice and use UHF references exclusively in the present work for energy calculations, but the question remains whether ROCCSD(T) or UCCSD(T) is preferable for the reference geometries of open-shell species.

\section{\label{sec:methods}Computational methods}

Most calculations in this paper were carried out using a development version of the CFOUR program system\cite{CFOUR} running on the CHEMFARM high-performance computing facility of the Faculty of Chemistry at Weizmann. Selected additional calculation were carried out using release versions of MOLPRO 2024\cite{MOLPRO} and MRCC 2024\cite{MRCC}.

The molecules considered are the 200-species W4-17 thermochemical benchmark\cite{jmlm273}. These span a range of inorganic and organic molecules, first-row and second-row (including `pseudohypervalent' species in which the $3d$ acts as an `honorary valence orbital'\cite[and references therein]{jmlm191,jmlm307}), and range from essentially purely dynamical correlation (such as H$_2$O and SiF$_4$) to strong static correlation (such as O$_3$, S$_4$, C$_2$, and BN). In the present work, we focus mostly on the W4-08\cite{jmlm215} subset of W4-11\cite{jmlm235} (and, in turn, W4-17).

Basis sets considered are the Dunning-Peterson `correlation consistent core-valence $n$-tuple zeta', cc-pCV$n$Z ($n$=T,Q,5,6), and their core-valence weighted variant, cc-pwCV$n$Z ($n$=T,Q,5), both described in Ref.\cite{pwCVnZ}, in combination with the standard cc-pVnZ basis sets\cite{Dunning1989,Woon1993} on hydrogen.  The original W4 theory employed aug-cc-pV(n+d)Z basis sets\cite{Dunning2001} on second-row atoms, as otherwise, major SCF-level errors are made (reaching 50 kcal/mol in \ce{HClO4}!)\cite{jmlm191} for species including a second-row atom in a high oxidation state. Effectively (see Ref.\cite{jmlm191} and references therein) an additional `tight' (high-exponent) $d$ function is needed in order to describe the $3d$ `honorary valence orbital'\cite{jmlm307} in such species, and hence its ability to accept back-bonding from chalcogen and halogen lone pairs. The core-valence basis sets, especially for higher cardinal number, already contain $d$ functions in the required high-exponent range, and therefore do not require `+d' additions except possibly\cite{Yockel2008} for the lowest cardinal numbers.

Karton\cite{Karton2022b} investigated the effect of such basis functions on higher-order correlation effects and concluded it to be negligible.\cite{Karton2022b} However, presently we will consider cc-pwCV$n$Z basis sets for these contributions anyhow, which bypasses the possible deficiency.

While monitoring some of the larger calculations required in this work, it was discovered that, especially in (Q) steps, CFOUR alternated bursts of OpenMP-parallel activity with stretches of single-core activity. The latter, upon more detailed analysis, were surprisingly revealed to be the weighting of the cluster amplitudes by the Fock denominators, a subleading-order step in computational time complexity. These denominators were calculated "on-the-fly", with the goal of easing implementation of perturbation theories with alternative zeroth-order Hamiltonians. The denominator weighting was reimplemented using pre-computed blocks of virtual orbitals for each irreducible representation in the typical direct-product decomposition order, parallelized at the granularity of each individual occupied index combination, and accelerated with AVX2 and NEON vector intrinsics for the x86\textunderscore64 and aarch64 architectures, respectively. This approach eliminated the above bottleneck at the cost of minimal additional memory consumption, leading to wall-time speedups by factors as high as 10-30 in larger cases.

\subsection{A remark concerning reference geometries}

Starting geometries were taken from the ESI of the W4-17 paper\cite{jmlm273}. For closed-shell species, geometries were optimized using MOLPRO at both the frozen-core CCSD(T)/cc-pV(Q+d)Z level and the active-core CCSD(T)/cc-pwCVQZ level (in which only the very deep $1s$ cores on Al--Cl were frozen). For open-shell species, we additionally considered both UHF and ROHF reference variants of each, to wit, UCCSD(T)\cite{Rag89} and ROCCSD(T)\cite{Wat93}. Geometric parameters were converged to five decimal places RMS; analytical derivatives were used to the extent possible. The reoptimized geometries are provided in the ESI (electronic supporting information).

For nearly all W4-17 species, the (RO)CCSD(T)/cc-pV(Q+d)Z reoptimized geometries agree with the initial geometries within the uncertainty of the original optimizations (which at the time they were carried out, 15-20 years ago, were incomparably more strenuous on available hardware). Nontrivial discrepancies were found for the following closed-shell species: BN, \ce{C2}, \ce{CF2},  \ce{CH3F}, \ce{SiH3F}, \ce{F2CO}, \ce{HOClO}. A handful of open-shell species agree well at the UCCSD(T) level but less so at the ROCCSD(T) level: \ce{B2}, \ce{CN}, \ce{S2}, \ce{SSH}, \ce{H2CCN}. 

This raises the question as to which is actually preferred for open-shell geometry optimizations: UCCSD(T) or ROCCSD(T)? We shall discuss this momentarily.

\section{\label{sec:results}Results and Discussion}

\subsection{ROCCSD(T) vs. UCCSD(T) geometries}

For most open-shell W4-17 systems, we find that ROCCSD(T)/cc-pV(Q+d)Z and UCCSD(T)/cc-pV(Q+d)Z geometries differ in just the fourth decimal place. 

However, more significant differences are found for a handful of species with strong spin contamination: in the W4-08 subset they are \{CN, CCH, \ce{CH2CH}, \ce{H2CN}\} with $<\hat{S}^2>$=\{1.15, 1.15, 0.97, 0.96\}. (For the set complement W4-17 $\setminus$ W4-08, they are allyl \ce{CH2=CH-CH2} and \ce{H2CCN} with $<\hat{S}^2>$=0.96 and 0.95, respectively.)

The best illustration is probably given by the CN and CCH radicals. As seen in Table \ref{tab:ROCCSD}, 
fully iterative ROCCSDT and UCCSDT bond distances for CN differ by just 0.0001 \AA, an order of magnitude less than between ROCCSD and UCCSD. The latter however pales in comparison to the 0.005 \AA\ between ROCCSD(T) and UCCSD(T). Of these two, ROCCSD(T) is much closer to CCSDT than is UCCSD(T). Admittedly, CCSDT is short about 0.0025 \AA\ owing to the neglect of connected quadruples. Since UCCSD(T) errs in the opposite direction, however, it falls even further short of the CCSDTQ(5)$_\Lambda$ result than it does of the CCSDT value.

Turning now to CCH radical, we see a 0.0035 \AA\ difference in C$\equiv$C bond distance between ROCCSD(T) and UCCSD(T), but of just 0.0004 \AA\ for $r_{\rm CH}$. Once again, ROCCSDT and UCCSDT are basically interchangeable (difference of just 0.0001 \AA\ for $r_{\rm CC}$, even less for $r_{\rm CH}$). ROCCSD and UCCSD differ by about an order of magnitude more, but again the difference is way less significant than for the (T) methods, and ROCCSD(T) is much closer to \{R,U\}CCSDT. 

For vinyl radical, the same observations are repeated.

Thus it would seem clear that ROCCSD(T) is to be preferred over UCCSD(T) for highly spin-contaminated cases. This choice also satisfies the `above all, do no harm' test, since in radicals with little spin contamination, we found that ROCCSD(T) and UCCSD(T) yield almost interchangeable geometries.

\begin{table}[h]

\caption{Comparison between ROCCSD(T), UCCSD(T), CCSDT, and higher level bond lengths (\AA) for three radicals prone to strong spin contamination\label{tab:ROCCSD}.}
\begin{tabular}{lrrrr}
\hline\hline
&\multicolumn{2}{c}{CN($^2\Sigma^+$) cc-pVTZ}&\multicolumn{2}{c}{CCH($^2\Sigma^+$) cc-pVDZ}\\
	   & $r_\mathrm{CN}$	& w.r.t CCSDT& $r_\mathrm{CC}$ & $r_\mathrm{CH}$\\
       \hline
ROCCSD	    &1.16884 & -0.00933& 1.22884&   1.07849\\
UCCSD    	&1.16773 & -0.01045&1.22795 &  1.07823    \\
ROCCSD(T)	&1.17929	& 0.00111 & 1.23536 &   1.08016\\
UCCSD(T)	&1.17451	& -0.00367 & 1.23183 &   1.07974\\
ROCCSDT	    &1.17829    & 0.00011 & 1.23538  &  1.08008\\
UCCSDT     	&1.17818	& REFERENCE & 1.23526 &  1.08004\\
UCCSDT(Q)	&1.18097	 & 0.00279 \\
UCCSDT(Q)$_\Lambda$	&1.18060  & 0.00242 \\
UCCSDTQ	  & 1.18038 & 0.00220 \\
UCCSDTQ(5)$_\Lambda$	&1.18071 	& 0.00253 \\
\hline
&\multicolumn{4}{c}{\ce{CH2CH}($^2A'$) cc-pVDZ}\\
     & $r_\mathrm{CH1}$	&   $r_\mathrm{CC}$	 &    $r_\mathrm{CH2}$	&   $r_\mathrm{CH3}$\\
     \hline
ROCCSD	   & 1.09475	& 1.33046 &  1.09833 & 1.10347\\
UCCSD	   & 1.09477 & 1.32948 &  1.09827 & 1.10341\\
ROCCSD(T) & 1.09635	& 1.33528 &  1.09982 & 1.10534\\
UCCSD(T)   & 1.09626	& 1.33264 & 1.09973	& 1.10522\\

ROCCSDT	  &  1.09639	& 1.33560 & 1.09982	& 1.10544\\
UCCSDT	  &  1.09646	& 1.33567 & 1.09988	& 1.10547\\
  \hline\hline
\end{tabular}
\end{table}

\subsection{Effect of CV geometry shift on thermochemistry}

As first reported in 1995,\cite{jmlm067} the core-valence optimized geometry invariably features shorter bond distances (see ESI). The contractions range from about 0.001--0.002 \AA\ for a C-H bond via 0.003--0.004 \AA\ for CC bonds and 0.008 \AA\ for the BB bond distance in diborane to 0.009--0.013 \AA\ in second-row species such as \ce{P4} and \ce{S4}. 

\begin{figure*}[t]
  \centering
  \includegraphics[width=0.48\textwidth]{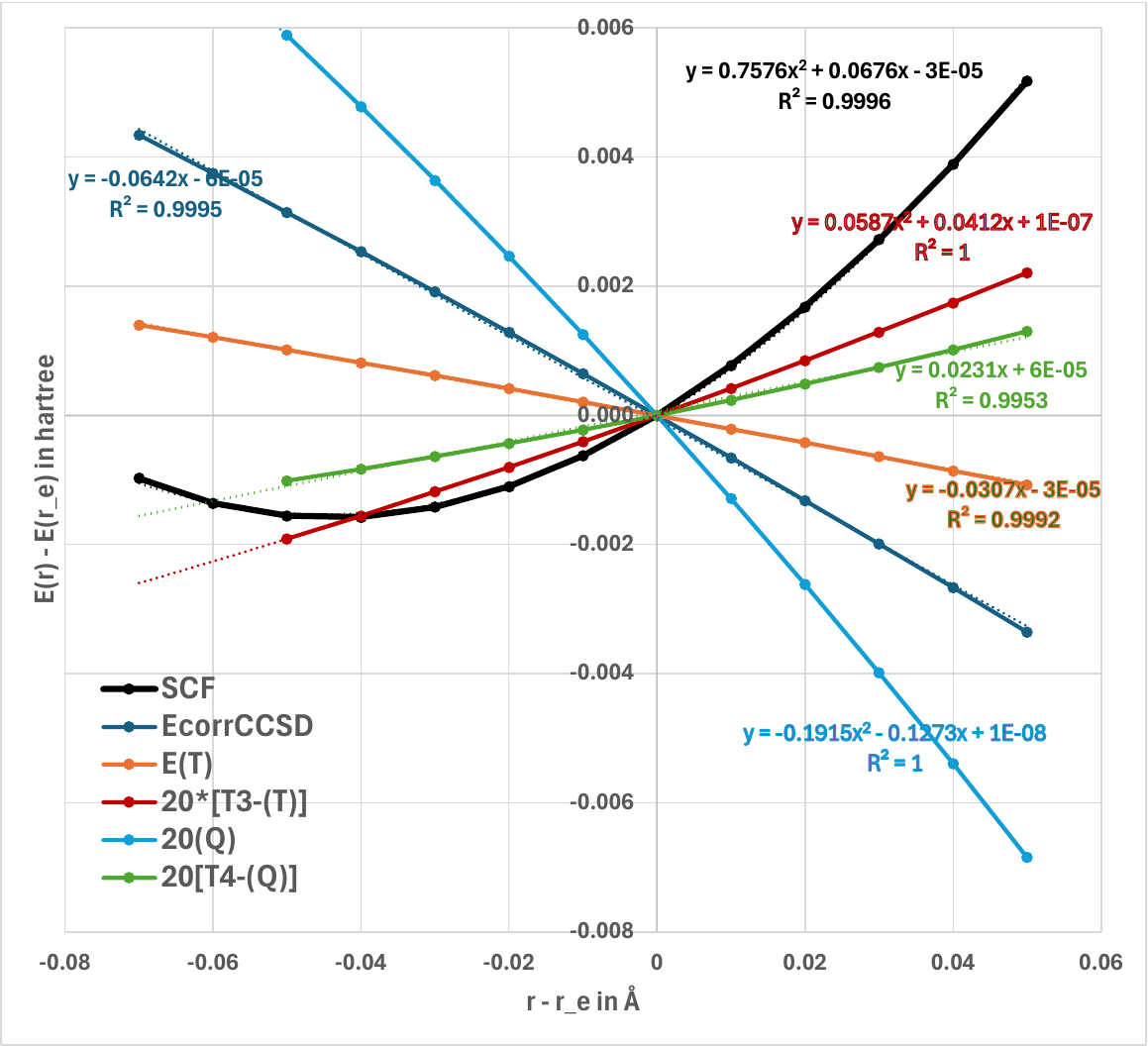}\hfill
  \includegraphics[width=0.48\textwidth]{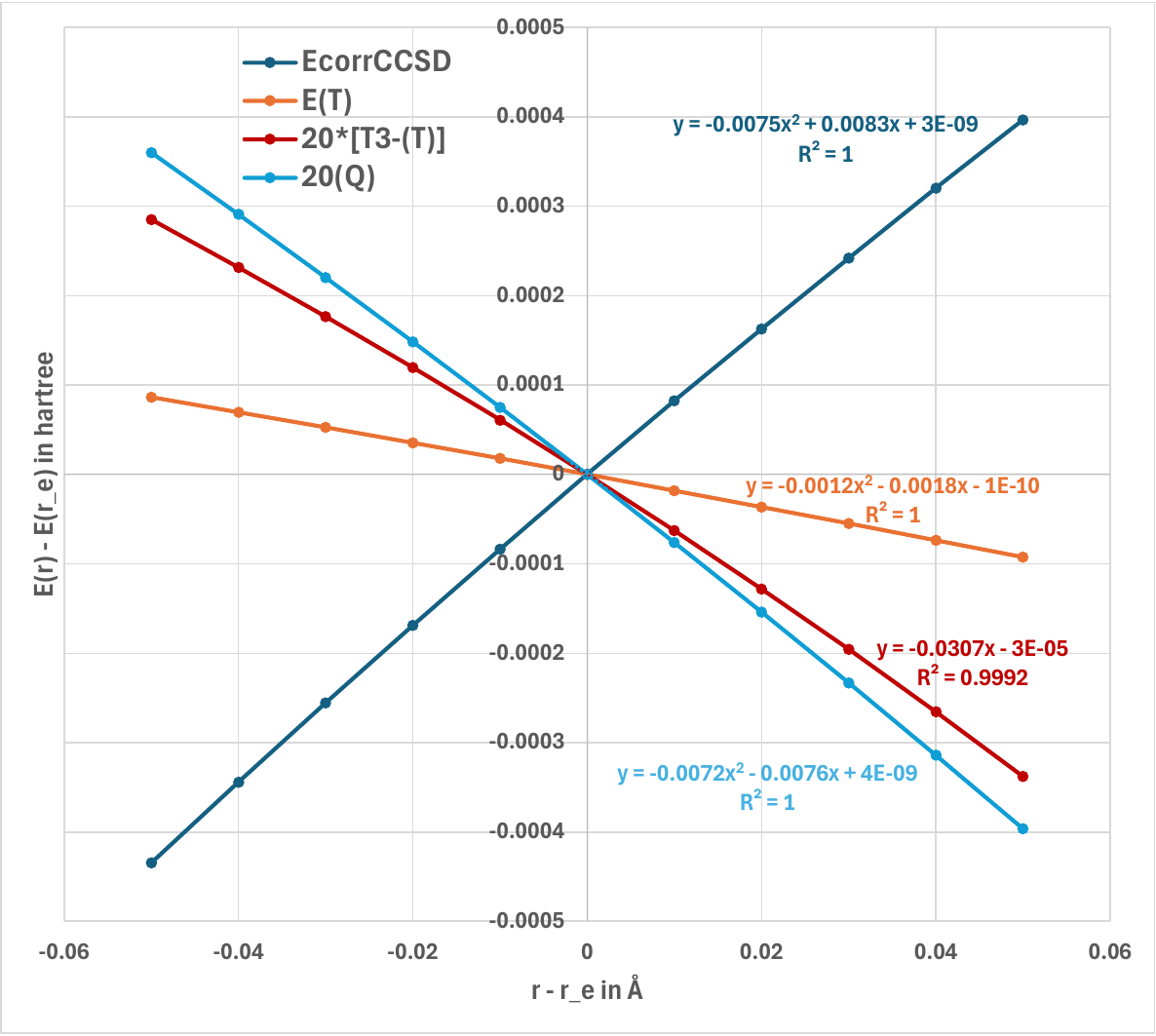}
  \caption{For the \ce{P2} diatomic in the cc-pwCVTZ basis set; (a) Left-hand pane: dependence of different valence energy components (hartree) on the displacement (\AA) from the  CCSD(T)/pwCVQZ reference bond distance $r_e$; (b) Right-hand pane: same graph for the subvalence contributions.}
  \label{fig:P2_val_cv}
\end{figure*}

Some light on this may be shed by Figure~\ref{fig:P2_val_cv}, a plot of the SCF and different correlation energy components for a representative diatomic, \ce{P2}. (We note that stretching curves near $r_e$ for other diatomics such as CO are qualitatively similar: examples are given in the Supporting Information.) It is remarkable, incidentally, how close to linear the correlation components are
in the displacement from equilibrium $r - r_e$. The minimum of a potential $E(r)=k(r-r_e)^2/2+C(r-r_e)$, where $k$ is a stretching force constant and C a constant slope, will be given by $r_{\rm min}=r_e - C$. Thus, as seen in the left-hand panel of the figure for valence correlation: a positive slope, such as seen for $T_3 - ( T)$ valence, will shorten the bond, and a negative slope, such as seen for valence CCSD correlation, (T), and (Q), will lengthen it. In the right-hand panel, we see that CCSD inner-shell correlation has a strong positive slope, which explains the observed bond contraction upon introducing inner-shell correlation. Higher-order correlation effects all temper this tendency.

\begin{table}[]
\caption{Breakdown by components of geometry shift effects on TAE$_0$ (kcal/mol). A positive number means TAE$_0$@CCSD(T)/pwCVQZ $>$ TAE$_0$@CCSD(T)/pV(Q+d)Z. \label{tab:geomshift}}

{

\fontsize{6}{7}\selectfont
\begin{tabular}{llllllll}
\hline\hline
 &
  \multicolumn{1}{c}{\textbf{ACV\{5,6\}Z}} &
  \multicolumn{1}{c}{\textbf{ACV\{5,6\}Z}} &
  \multicolumn{1}{c}{\textbf{pwCVTZ}} &
  \multicolumn{1}{c}{\textbf{pwCVTZ}} &
  \multicolumn{1}{c}{\textbf{pwCVTZ}} &
  \multicolumn{1}{c}{\textbf{pwCVTZ}} &
  \multicolumn{1}{c}{\textbf{pwCVTZ}} \\
 &

  \multicolumn{1}{c}{\textbf{CCSD(T) val}} &
  \multicolumn{1}{c}{\textbf{CCSD(T) CV}} &
  \multicolumn{1}{c}{\textbf{$T_3$-(T) val}} &
  \multicolumn{1}{c}{\textbf{$\Delta$(Q) val}} &
  \multicolumn{1}{c}{{\textbf{CV $T_3$-(T) }}} &
  \multicolumn{1}{c}{{\textbf{CV (Q)}}} &
  \multicolumn{1}{c}{{\textbf{(Q)$_\Lambda$ - (Q) val }}} \\
  \hline
  \ce{B2H6}   & 0.005   & \cellcolor[HTML]{D8EEE0}0.042 & \cellcolor[HTML]{FBF5F8}-0.003 & \cellcolor[HTML]{FBEDF0}-0.001 & 0.000  & 0.000   & 0.000   \\
\ce{BHF2}   & 0.006   & \cellcolor[HTML]{E2F2E9}0.030 & \cellcolor[HTML]{FBF9FC}-0.001 & \cellcolor[HTML]{FBEBEE}-0.002 & 0.000  & 0.000   & 0.000   \\
\ce{BF3}    & 0.011   & \cellcolor[HTML]{DBEFE3}0.038 & \cellcolor[HTML]{FBF9FC}-0.001 & \cellcolor[HTML]{FBEAED}-0.004 & 0.000  & 0.000   & 0.000   \\
\ce{C2H6}   & 0.018   & \cellcolor[HTML]{EAF5EF}0.021 & \cellcolor[HTML]{FBFAFD}-0.001 & \cellcolor[HTML]{FBEDEF}-0.001 & 0.000  & 0.000   & 0.000   \\
\ce{H2CN}   & 0.029   & \cellcolor[HTML]{EBF5F0}0.020 & \cellcolor[HTML]{FAD6D9}-0.014 & \cellcolor[HTML]{FBE9EC}-0.005 & -0.001 & 0.000   & 0.002   \\
\ce{NCCN}   & 0.031   & \cellcolor[HTML]{CFEAD8}0.052 & \cellcolor[HTML]{A0BBDF}0.012  & \cellcolor[HTML]{FAD2D5}-0.030 & -0.001 & 0.000   & 0.002   \\
\ce{CH2NH2} & 0.028   & \cellcolor[HTML]{EAF5F0}0.021 & \cellcolor[HTML]{FBF7FA}-0.002 & \cellcolor[HTML]{FBECEF}-0.002 & 0.000  & 0.000   & 0.000   \\
\ce{CH3NH}  & 0.020   & \cellcolor[HTML]{EEF6F3}0.017 & \cellcolor[HTML]{FBF9FC}-0.001 & \cellcolor[HTML]{FBECEF}-0.002 & 0.000  & 0.000   & 0.000   \\
\ce{CH3NH2} & 0.023   & \cellcolor[HTML]{ECF6F1}0.019 & \cellcolor[HTML]{FBFBFE}0.000  & \cellcolor[HTML]{FBECEE}-0.002 & 0.000  & 0.000   & 0.000   \\
\ce{CF2}    & 0.012   & \cellcolor[HTML]{EDF6F2}0.018 & \cellcolor[HTML]{FBF5F8}-0.003 & \cellcolor[HTML]{FBE9EC}-0.005 & 0.000  & 0.000   & 0.001   \\
\ce{N2H}    & 0.038   & \cellcolor[HTML]{EFF7F4}0.015 & \cellcolor[HTML]{FAD2D5}-0.016 & \cellcolor[HTML]{FBE7EA}-0.007 & -0.001 & 0.000   & 0.003   \\
\ce{t-N2H2} & 0.026   & \cellcolor[HTML]{EFF7F4}0.015 & \cellcolor[HTML]{E3EAF6}0.003  & \cellcolor[HTML]{FBE5E8}-0.010 & 0.000  & 0.000   & 0.001   \\
\ce{N2H4}   & 0.029   & \cellcolor[HTML]{ECF6F1}0.019 & \cellcolor[HTML]{F8FAFE}0.001  & \cellcolor[HTML]{FBEAEC}-0.005 & 0.000  & 0.000   & 0.000   \\

\ce{FOOF}   & 0.027   & \cellcolor[HTML]{F4F9F8}0.009 & \cellcolor[HTML]{FBF5F8}-0.003 & \cellcolor[HTML]{FAC6C8}-0.043 & -0.001 & -0.001  & 0.008   \\
\ce{AlF3}   & N/A & N/A                       & \cellcolor[HTML]{FBFBFF}0.000  & \cellcolor[HTML]{FBE7E9}-0.008 & -0.001 & -0.001  & 0.001   \\
\ce{Si2H6}  & -0.024  & \cellcolor[HTML]{A4D9B3}0.101 & \cellcolor[HTML]{FBF2F5}-0.004 & \cellcolor[HTML]{FBEDEF}-0.001 & -0.001 & 0.000   & 0.000   \\
\ce{P4}     & 0.059   & \cellcolor[HTML]{94D2A5}0.120 & \cellcolor[HTML]{A2BDE0}0.012  & \cellcolor[HTML]{FAC4C6}-0.045 & -0.002 & -0.003  & 0.004   \\
\ce{SO2}    & 0.037   & \cellcolor[HTML]{C5E6D0}0.063 & \cellcolor[HTML]{CDDBEF}0.006  & \cellcolor[HTML]{FAC2C5}-0.047 & -0.001 & -0.002  & 0.007   \\
\ce{SO3}    & 0.059   & \cellcolor[HTML]{9FD7AF}0.107 & \cellcolor[HTML]{B7CCE7}0.009  & \cellcolor[HTML]{FABBBD}-0.055 & -0.002 & -0.002  & 0.006   \\
\ce{OCS}    & 0.019   & \cellcolor[HTML]{D9EEE1}0.041 & \cellcolor[HTML]{FAFAFE}0.000  & \cellcolor[HTML]{FBE0E3}-0.014 & 0.000  & -0.001  & 0.002   \\
\ce{CS2}    & 0.017   & \cellcolor[HTML]{CDE9D6}0.054 & \cellcolor[HTML]{F4F7FD}0.001  & \cellcolor[HTML]{FBDCDF}-0.019 & 0.000  & -0.002  & 0.002   \\
\ce{S2O}    & 0.048   & \cellcolor[HTML]{C8E7D2}0.060 & \cellcolor[HTML]{DEE7F5}0.004  & \cellcolor[HTML]{FABBBE}-0.054 & -0.002 & -0.003  & 0.010   \\
\ce{S3}     & 0.057   & \cellcolor[HTML]{C7E7D2}0.061 & \cellcolor[HTML]{FBF7FA}-0.002 & \cellcolor[HTML]{FABABD}-0.056 & -0.002 & -0.005  & 0.012   \\
\ce{S4 (C_{2v})}      & 0.080   & \cellcolor[HTML]{B1DEBE}0.087 & \cellcolor[HTML]{5A8AC6}0.022  & \cellcolor[HTML]{F8696B}-0.144 & -0.009 & -0.003  & 0.035   \\
\ce{CCl2}   & 0.026   & \cellcolor[HTML]{DDF0E4}0.036 & \cellcolor[HTML]{FBE8EB}-0.007 & \cellcolor[HTML]{FBE5E7}-0.010 & 0.000  & -0.001  & 0.003   \\
\ce{AlCl3}  & -0.020  & \cellcolor[HTML]{63BE7B}0.175 & \cellcolor[HTML]{FBF3F5}-0.003 & \cellcolor[HTML]{FBEAEC}-0.005 & -0.001 & -0.001  & 0.001   \\
\ce{ClCN}   & 0.020   & \cellcolor[HTML]{DDF0E4}0.036 & \cellcolor[HTML]{D9E3F3}0.005  & \cellcolor[HTML]{FBDFE2}-0.016 & -0.001 & 0.000   & 0.001   \\
\ce{OClO}   & 0.047   & \cellcolor[HTML]{DBEFE2}0.039 & \cellcolor[HTML]{FBF7FA}-0.002 & \cellcolor[HTML]{FABDC0}-0.052 & 0.000  & -0.002  & 0.010   \\

\ce{Cl2O}   & 0.033   & \cellcolor[HTML]{EDF6F2}0.018 & \cellcolor[HTML]{F5F8FD}0.001  & \cellcolor[HTML]{FBDDE0}-0.018 & -0.001 & -0.001  & 0.002   \\
\ce{BN (^3\Pi)}     & 0.018   & \cellcolor[HTML]{E0F1E7}0.032 & \cellcolor[HTML]{FBDADD}-0.013 & \cellcolor[HTML]{FBE1E4}-0.014 & 0.000  & -0.001  & 0.005   \\
\ce{CF}     & 0.006   & \cellcolor[HTML]{F3F9F7}0.011 & \cellcolor[HTML]{FBF5F8}-0.002 & \cellcolor[HTML]{FBEBEE}-0.003 & 0.000  & 0.000   & 0.001   \\
\ce{CH2C}   & 0.012   & \cellcolor[HTML]{E6F3EC}0.025 & \cellcolor[HTML]{FBFAFD}-0.001 & \cellcolor[HTML]{FBEAED}-0.004 & 0.000  & 0.000   & 0.001   \\
\ce{CH2CH}  & 0.027   & \cellcolor[HTML]{E8F4EE}0.023 & \cellcolor[HTML]{FAD6D8}-0.014 & \cellcolor[HTML]{FBE9EC}-0.005 & -0.001 & 0.000   & 0.001   \\
\ce{C2H4}   & 0.014   & \cellcolor[HTML]{E9F4EE}0.023 & \cellcolor[HTML]{FAFBFF}0.000  & \cellcolor[HTML]{FBEAED}-0.004 & 0.000  & 0.000   & 0.000   \\
\ce{CH2NH}  & 0.018   & \cellcolor[HTML]{ECF6F1}0.019 & \cellcolor[HTML]{F1F5FC}0.001  & \cellcolor[HTML]{FBE8EB}-0.006 & 0.000  & 0.000   & 0.000   \\
\ce{HCO}    & 0.015   & \cellcolor[HTML]{EEF7F3}0.017 & \cellcolor[HTML]{FBF8FB}-0.001 & \cellcolor[HTML]{FBE7EA}-0.007 & 0.000  & 0.000   & 0.001   \\
\ce{H2CO}   & 0.014   & \cellcolor[HTML]{EFF7F4}0.016 & \cellcolor[HTML]{F8F9FE}0.001  & \cellcolor[HTML]{FBE9EB}-0.006 & 0.000  & 0.000   & 0.001   \\
\ce{CO2}    & 0.023   & \cellcolor[HTML]{E4F2EA}0.028 & \cellcolor[HTML]{E9EFF9}0.003  & \cellcolor[HTML]{FBE1E4}-0.013 & 0.000  & 0.000   & 0.001   \\
\ce{HNO}    & 0.020   & \cellcolor[HTML]{F3F9F7}0.011 & \cellcolor[HTML]{E3EAF6}0.003  & \cellcolor[HTML]{FBE4E7}-0.010 & 0.000  & 0.000   & 0.001   \\
\ce{NO2}    & 0.029   & \cellcolor[HTML]{EBF5F1}0.020 & \cellcolor[HTML]{F2F5FC}0.001  & \cellcolor[HTML]{FAD5D8}-0.026 & -0.001 & -0.001  & 0.004   \\
\ce{N2O}    & 0.029   & \cellcolor[HTML]{E6F4EC}0.025 & \cellcolor[HTML]{B4C9E6}0.010  & \cellcolor[HTML]{FAD3D6}-0.029 & -0.001 & 0.000   & 0.004   \\
\ce{O3}     & 0.029   & \cellcolor[HTML]{EFF7F4}0.015 & \cellcolor[HTML]{B8CCE7}0.009  & \cellcolor[HTML]{FAB3B5}-0.064 & -0.001 & -0.001  & 0.013   \\
\ce{HOO}    & 0.017   & \cellcolor[HTML]{F7FAFB}0.006 & \cellcolor[HTML]{FBEBEE}-0.006 & \cellcolor[HTML]{FBE8EB}-0.006 & 0.000  & 0.000   & 0.002   \\
\ce{HOOH}   & 0.017   & \cellcolor[HTML]{F7FAFB}0.006 & \cellcolor[HTML]{F5F7FD}0.001  & \cellcolor[HTML]{FBE7EA}-0.007 & 0.000  & 0.000   & 0.000   \\
\ce{F2O}    & 0.019   & \cellcolor[HTML]{F8FBFB}0.005 & \cellcolor[HTML]{FBFBFF}0.000  & \cellcolor[HTML]{FBE2E4}-0.013 & 0.000  & 0.000   & 0.002   \\
\ce{HOCl}   & 0.020   & \cellcolor[HTML]{F5F9F9}0.009 & \cellcolor[HTML]{FBFBFE}0.000  & \cellcolor[HTML]{FBE8EA}-0.007 & 0.000  & 0.000   & 0.001   \\
\ce{SSH}    & 0.027   & \cellcolor[HTML]{E6F3EC}0.026 & \cellcolor[HTML]{FBE9EC}-0.007 & \cellcolor[HTML]{FBE9EC}-0.005 & 0.000  & -0.001  & 0.001   \\
\ce{B2 ($^3\Sigma^-_g$)}      & -0.002  & \cellcolor[HTML]{E8F4EE}0.023 & \cellcolor[HTML]{FBDCDF}-0.012 & \cellcolor[HTML]{FBEDF0}-0.001 & 0.000  & 0.000   & 0.000   \\
\ce{BH}     & 0.002   & \cellcolor[HTML]{F8FBFC}0.005 & \cellcolor[HTML]{FBFBFE}0.000  & \cellcolor[HTML]{FBEEF1}0.000  & 0.000  & 0.000   & 0.000   \\
\ce{BH3}    & 0.004   & \cellcolor[HTML]{F1F8F6}0.013 & \cellcolor[HTML]{FBFAFD}-0.001 & \cellcolor[HTML]{FBEEF1}0.000  & 0.000  & 0.000   & 0.000   \\
\ce{BN ($^1\Sigma^+$)}      & 0.004   & \cellcolor[HTML]{E4F2EA}0.028 & \cellcolor[HTML]{F8696B}-0.056 & \cellcolor[HTML]{FCFCFF}0.015  & 0.000  & 0.002   & -0.036  \\
\ce{BF}     & 0.000   & \cellcolor[HTML]{E8F4EE}0.024 & \cellcolor[HTML]{FBFAFD}-0.001 & \cellcolor[HTML]{FBECEF}-0.002 & 0.000  & 0.000   & 0.001   \\
\ce{NH ($^3\Sigma^-$)}      & 0.005   & \cellcolor[HTML]{FBFCFF}0.001 & \cellcolor[HTML]{FBFBFE}0.000  & \cellcolor[HTML]{FBEEF1}0.000  & 0.000  & 0.000   & 0.000   \\
\ce{NH2}    & 0.010   & \cellcolor[HTML]{FAFBFD}0.003 & \cellcolor[HTML]{FBFAFD}0.000  & \cellcolor[HTML]{FBEDF0}0.000  & 0.000  & 0.000   & 0.000   \\
\ce{HCN}    & 0.018   & \cellcolor[HTML]{E9F5EF}0.022 & \cellcolor[HTML]{DAE4F3}0.005  & \cellcolor[HTML]{FBE2E5}-0.012 & 0.000  & 0.000   & 0.001   \\
\ce{HOF}    & 0.012   & \cellcolor[HTML]{F9FBFD}0.004 & \cellcolor[HTML]{F9FAFE}0.000  & \cellcolor[HTML]{FBE9EC}-0.005 & 0.000  & 0.000   & 0.001   \\
\ce{AlH}    & -0.001  & \cellcolor[HTML]{F7FAFB}0.007 & \cellcolor[HTML]{FBFAFD}-0.001 & \cellcolor[HTML]{FBEEF1}0.000  & 0.000  & 0.000   & 0.000   \\
\ce{AlH3}   & -0.019  & \cellcolor[HTML]{CBE9D5}0.056 & \cellcolor[HTML]{FBF6F9}-0.002 & \cellcolor[HTML]{FBEDF0}0.000  & 0.000  & 0.000   & 0.000   \\
\ce{AlF}    & -0.011  & \cellcolor[HTML]{D6EDDE}0.044 & \cellcolor[HTML]{FBFAFD}-0.001 & \cellcolor[HTML]{FBECEF}-0.002 & 0.000  & 0.000   & 0.001   \\
\ce{AlCl}   & 0.002   & \cellcolor[HTML]{E6F3EC}0.026 & \cellcolor[HTML]{FBF9FC}-0.001 & \cellcolor[HTML]{FBEDF0}-0.001 & 0.000  & 0.000   & 0.000   \\
\ce{SiH}    & 0.001   & \cellcolor[HTML]{F8FBFC}0.005 & \cellcolor[HTML]{FBFAFD}-0.001 & \cellcolor[HTML]{FBEEF1}0.000  & 0.000  & 0.000   & 0.000   \\
\ce{SiH4}   & -0.009  & \cellcolor[HTML]{D8EEE0}0.042 & \cellcolor[HTML]{FBF6F9}-0.002 & \cellcolor[HTML]{FBEDF0}0.000  & 0.000  & 0.000   & 0.000   \\
\ce{SiO}    & 0.006   & \cellcolor[HTML]{CAE8D4}0.057 & \cellcolor[HTML]{E9EFF9}0.003  & \cellcolor[HTML]{FBD7DA}-0.024 & -0.001 & -0.001  & 0.006   \\
\ce{SiF}    & 0.003   & \cellcolor[HTML]{E6F3EC}0.026 & \cellcolor[HTML]{FBF8FB}-0.001 & \cellcolor[HTML]{FBEBEE}-0.003 & 0.000  & 0.000   & 0.001   \\
\ce{CS}     & 0.015   & \cellcolor[HTML]{E0F1E7}0.033 & \cellcolor[HTML]{FBFBFE}0.000  & \cellcolor[HTML]{FBE3E6}-0.012 & 0.000  & -0.001  & 0.002   \\
\ce{H2}     & 0.002   & \cellcolor[HTML]{FCFCFF}0.000 & \cellcolor[HTML]{FCFCFF}0.000  & \cellcolor[HTML]{FBEEF1}0.000  & 0.000  & 0.000   & 0.000   \\
\ce{OH}     & 0.005   & \cellcolor[HTML]{FBFCFE}0.002 & \cellcolor[HTML]{FBFBFE}0.000  & \cellcolor[HTML]{FBEEF1}0.000  & 0.000  & 0.000   & 0.000   \\
\ce{HF}     & 0.005   & \cellcolor[HTML]{FCFCFF}0.000 & \cellcolor[HTML]{FCFCFF}0.000  & \cellcolor[HTML]{FBEEF1}0.000  & 0.000  & 0.000   & 0.000   \\
\ce{H2O}    & 0.009   & \cellcolor[HTML]{FBFCFE}0.002 & \cellcolor[HTML]{FCFCFF}0.000  & \cellcolor[HTML]{FBEDF0}-0.001 & 0.000  & 0.000   & 0.000   \\
\ce{CH (^2\Pi)}     & 0.004   & \cellcolor[HTML]{FAFCFE}0.002 & \cellcolor[HTML]{FBFBFE}0.000  & \cellcolor[HTML]{FBEEF1}0.000  & 0.000  & 0.000   & 0.000   \\
\ce{CH2 (^3B1)}& 0.005   & \cellcolor[HTML]{F8FBFC}0.005 & \cellcolor[HTML]{FBFAFD}-0.001 & \cellcolor[HTML]{FBEEF1}0.000  & 0.000  & 0.000   & 0.000   \\
\ce{CH3}    & 0.008   & \cellcolor[HTML]{F7FAFB}0.006 & \cellcolor[HTML]{FBFAFD}-0.001 & \cellcolor[HTML]{FBEEF0}0.000  & 0.000  & 0.000   & 0.000   \\
\ce{CH4}    & 0.010   & \cellcolor[HTML]{F5F9F9}0.009 & \cellcolor[HTML]{FBFAFD}0.000  & \cellcolor[HTML]{FBEEF0}0.000  & 0.000  & 0.000   & 0.000   \\
\ce{CCH}    & 0.037   & \cellcolor[HTML]{E7F4ED}0.025 & \cellcolor[HTML]{FAC2C4}-0.022 & \cellcolor[HTML]{FBE5E8}-0.009 & -0.001 & 0.000   & 0.002   \\
\ce{C2H2}   & 0.015   & \cellcolor[HTML]{E7F4ED}0.025 & \cellcolor[HTML]{E3EAF6}0.003  & \cellcolor[HTML]{FBE6E8}-0.009 & 0.000  & 0.000   & 0.001   \\
\ce{NH3}    & 0.015   & \cellcolor[HTML]{F7FAFB}0.006 & \cellcolor[HTML]{FBFBFE}0.000  & \cellcolor[HTML]{FBEDF0}-0.001 & 0.000  & 0.000   & 0.000   \\
\ce{C2}     & 0.010   & \cellcolor[HTML]{EAF5EF}0.022 & \cellcolor[HTML]{F7F8FD}0.001  & \cellcolor[HTML]{FBE0E2}-0.015 & 0.000  & 0.000   & 0.001   \\
\ce{N2}     & 0.022   & \cellcolor[HTML]{EDF6F2}0.018 & \cellcolor[HTML]{D7E2F2}0.005  & \cellcolor[HTML]{FBE1E4}-0.013 & 0.000  & 0.000   & 0.001   \\
\ce{CO}     & 0.013   & \cellcolor[HTML]{EBF5F1}0.020 & \cellcolor[HTML]{F6F8FD}0.001  & \cellcolor[HTML]{FBE7EA}-0.007 & 0.000  & 0.000   & 0.001   \\
\ce{CN}     & 0.049   & \cellcolor[HTML]{E9F5EF}0.022 & \cellcolor[HTML]{FAB4B7}-0.027 & \cellcolor[HTML]{FBDBDE}-0.020 & -0.002 & 0.000   & 0.003   \\
\ce{NO}     & 0.036   & \cellcolor[HTML]{F3F9F7}0.011 & \cellcolor[HTML]{FACED1}-0.017 & \cellcolor[HTML]{FBE4E7}-0.011 & 0.000  & 0.000   & 0.003   \\
\ce{O2}     & 0.013   & \cellcolor[HTML]{F6FAFA}0.008 & \cellcolor[HTML]{EFF3FB}0.002  & \cellcolor[HTML]{FBE2E5}-0.012 & 0.000  & 0.000   & 0.001   \\
\ce{OF}     & 0.012   & \cellcolor[HTML]{FAFBFD}0.003 & \cellcolor[HTML]{FBE9EC}-0.007 & \cellcolor[HTML]{FBE9EC}-0.005 & 0.000  & 0.000   & 0.002   \\
\ce{F2}     & 0.008   & \cellcolor[HTML]{FBFCFE}0.002 & \cellcolor[HTML]{F9FAFE}0.000  & \cellcolor[HTML]{FBE9EC}-0.005 & 0.000  & 0.000   & 0.001   \\
\ce{PH3}    & 0.003   & \cellcolor[HTML]{EDF6F2}0.018 & \cellcolor[HTML]{FBF8FB}-0.001 & \cellcolor[HTML]{FBEDF0}-0.001 & 0.000  & 0.000   & 0.000   \\
\ce{HS}     & 0.003   & \cellcolor[HTML]{F9FBFC}0.004 & \cellcolor[HTML]{FBFAFD}-0.001 & \cellcolor[HTML]{FBEDF0}0.000  & 0.000  & 0.000   & 0.000   \\
\ce{H2S}    & 0.004   & \cellcolor[HTML]{F5F9F9}0.008 & \cellcolor[HTML]{FBFAFD}-0.001 & \cellcolor[HTML]{FBEDF0}-0.001 & 0.000  & 0.000   & 0.000   \\
\ce{HCl}    & 0.003   & \cellcolor[HTML]{FAFBFD}0.003 & \cellcolor[HTML]{FBFBFE}0.000  & \cellcolor[HTML]{FBEDF0}0.000  & 0.000  & 0.000   & 0.000   \\
\ce{SO}     & 0.027   & \cellcolor[HTML]{E8F4EE}0.024 & \cellcolor[HTML]{FBFBFE}0.000  & \cellcolor[HTML]{FBDDE0}-0.018 & 0.000  & -0.001  & 0.001   \\
\ce{ClO}    & 0.027   & \cellcolor[HTML]{F2F8F7}0.012 & \cellcolor[HTML]{FBE1E4}-0.010 & \cellcolor[HTML]{FBE9EC}-0.005 & 0.000  & -0.001  & 0.002   \\
\ce{ClF}    & 0.014   & \cellcolor[HTML]{F7FAFB}0.006 & \cellcolor[HTML]{FBF9FC}-0.001 & \cellcolor[HTML]{FBE9EC}-0.005 & 0.000  & 0.000   & 0.001   \\
\ce{P2}     & 0.023   & \cellcolor[HTML]{D2EBDB}0.049 & \cellcolor[HTML]{B0C6E4}0.010  & \cellcolor[HTML]{FAD0D3}-0.032 & -0.002 & -0.002  & 0.004   \\
\ce{S2}     & 0.026   & \cellcolor[HTML]{E4F3EA}0.028 & \cellcolor[HTML]{FBFAFD}-0.001 & \cellcolor[HTML]{FBE2E5}-0.013 & 0.000  & -0.001  & 0.001   \\
\ce{Cl2}    & 0.025   & \cellcolor[HTML]{F1F8F6}0.013 & \cellcolor[HTML]{FBF6F9}-0.002 & \cellcolor[HTML]{FBEAED}-0.004 & 0.000  & 0.000   & 0.000  \\
\hline\hline
\end{tabular}
}
\end{table}

More detailed information for the W4-08 subset can be found in Table~\ref{tab:geomshift}. At the HF level, some TAEs increase by over 1 kcal/mol when switching from `W4'  to `HEAT' (i.e., CCSD(T)/cc-pwCVQZ) reference geometries. However, this is greatly mitigated by the opposite change in the valence correlation component. As a result, geometry TAE shifts at the valence CCSD(T)/ACV\{5,6\}Z level are much more modest, typically in the 0.01-0.02 kcal/mol range for first-row compounds, with larger outliers for species like CN (with its strong spin contamination) and \ce{O3} (with its strong static correlation). For Al and Si compounds, some contributions are actually \emph{negative}. 

CCSD(T) inner-shell contribution to the geometry shift are consistently positive, and generally more significant, crossing the 0.1 kcal/mol threshold for many second-row species: 0.18 kcal/mol for \ce{AlCl3}, 0.12 kcal/mol for \ce{P4}, 0.11 kcal/mol for \ce{SO3}.
Valence higher-order triples contributions are only modestly affected by the geometry shift, with the pathological BN singly as the outlier. In contrast, geometry shifts for the valence connected quadruples are almost universally negative, reaching a surprising -0.14 kcal/mol for \ce{S4}, and generally being in the -0.05 kcal/mol range for many second-row compounds. 

Effects on the valence CCSDT(Q)$_\Lambda$ - CCSDT(Q) difference are negligible in most cases, with BN and \ce{S4} being outliers at -0.036 and +0.035 kcal/mol, respectively. 

As for the core-valence post-CCSD(T) corrections, these are already small(ish) to begin with, and the geometry effect on them is basically negligible. Adding everything up, we see partial cancellation in some cases like \ce{S4}, and are left with \ce{P4} at 0.15 kcal/mol and \ce{SO3} at 0.12 kcal/mol as the `champions'.  (It should be noted that already in a 1999 paper on \ce{SiF4},\cite{jmlm126}, a `note added in proof' mentioned a 0.15 kcal/mol reference geometry shift effect, while  a 2007 paper\cite{jmlm207} concerned with  \ce{P2} and \ce{P4} reported 0.05 and 0.13 kcal/mol, respectively.)

\subsection{CCSD and (T) basis set extrapolation}

Especially for second-row compounds, the subvalence correlation energy may rival the valence correlation. However, it has been established for at least two decades (see, e.g., Refs.\cite{jmlm127,jmlm200}) that for the B--F and Al--Cl block, subvalence contributions to total atomization energies are about two orders of magnitude smaller than the corresponding valence contributions, and that they converge fairly rapidly with the basis set (see Ref.\cite{jmlm281} for a detailed analysis).\footnote{The alkali and alkali earth metals are a special case\cite{jmlm161,jmlm164,jmlm307} where the (n-1)p orbitals take on `honorary valence orbital' character.} As an aside, and as likewise shown in Ref.\cite{jmlm281}, the received wisdom that states core-valence correlation much outweigh core-core correlation contributions\cite{CCandCV} in thermochemistry --- to the point that subvalence correlation is commonly referred to by the synecdoche `core-valence correlation' --- is largely correct for first-row molecules, but no longer holds in systems with adjacent second-row atoms, like \ce{S4}, \ce{P4}, and \ce{SSO}. As a result, ACV\{Q,5\}Z and \emph{a fortiori} ACV\{5,6\}Z subvalence contributions to TAE are not only converged with the basis set but fairly insensitive to the details of the  basis set extrapolation procedure.

Furthermore, it has been known since at least Helgaker et al.\cite{Helgaker1997} that (T) converges more rapidly with the basis set than the CCSD correlation energy: for a detailed analysis specifically for the W4-08 subset, we refer the reader to Ref.\cite{jmlm301}

This leaves the valence CCSD component as the most crucial one, and great effort has been expended by many groups on strategies for its extrapolation (e.g., Refs.\cite{Klopper2001,Schwenke2005,Feller2006})

In the emerging `SuperHEAT' approach (e.g., \cite{Thorpe2021_HEAT,Thorpe2023}), Thorpe et al. took a page from the playbook of Ref.\cite{Feller2006} by averaging the `Schwartz  formula'\cite{Schwartz1962}, $E_L \approx E_\infty + A/(L+1/2)^4$ with the simple Helgaker formula\cite{Halkier1998a} $E_L \approx E_\infty + A/L^3$, given as they tend to converge to the basis set limit from opposite directions. As shown by Schwenke\cite{Schwenke2005}, all two-point extrapolations can be reduced to the form $E_\infty \approx E_L + A_L[E_L-E_{L-1}]$ where we term $A_L$ a `Schwenke coefficient'. (For conversion formulas between the common extrapolations formulas and this form, see Ref.\cite{jmlm280}.) For the \{5,6\} basis set pair, Helgaker and Schwartz formulas correspond to $A_6(\textrm{Helgaker})=1.3736$ and $A_6(\textrm{Schwartz})=1.0518$, the average of which being $A_6(\textrm{SuperHEAT})=1.2127$. In different forms, the latter is equivalent to $A/L^{3.2983}$ or $A/(L-0.4946)^3$.  

One `sanity check' would be to compare with explicitly correlated coupled cluster theory, particularly with the more rigorous CCSD(F12*) approach.\cite{Haettig2010} (In Ref.\cite{jmlm285} we effectively availed ourselves of this check in the opposite direction, using CCSD data with Ranasinghe-Petersson\cite{Ranasinghe2013} 6ZaPa and 7ZaPa basis sets and the extrapolation formulas given there. We were thus able to show that the basis set limit CCSD(F12*) converges to is fundamentally compatible with the orbital basis set limit --- as it ought to be --- while more approximate methods such as CCSD-F12b\cite{AdlerWerner2007_CCSDF12b,KniziaAdlerWerner2009_CCSDT_F12b} neglect terms that remain thermochemically significant even for quintuple zeta basis sets.)

In the present work, we carried out CCSD(F12*)/aug-cc-pV(6+d)Z calculations with UHF references, using the implementation\cite{Kallay2021} in MRCC 2025.\cite{MRCC} [The following auxiliary basis sets were used: aug-cc-pV5Z-JK\cite{Weigend2006RI-JK}, aug-cc-pV5Z-OptRI\cite{HillPeterson2012OptRI},  and H\"attig's unpublished cc-pV6Z-RI from the Turbomole\cite{Turbomole} library.] For the first-row subset of W4-08, the resulting CCSD(F12*) valence contributions to TAE (see ESI) are in remarkable agreement (RMS deviation 0.035 kcal/mol) to ACV\{5,6\}Z with $A_6(\textrm{SuperHEAT})=1.2127$; minimizing RMSD with respect to $A_6$ yields $A_6(\textrm{opt})=1.262$ for RMSD=0.025 kcal/mol. 

The latter is almost identical to Schwenke's $A_6(\textrm{AVnZ})=1.266$; in Ref.\cite{jmlm280}, Table 1, footnote b, one of us found $A_6$=1.283 by fitting against 12 basis set limit CCSD-R12 energies from Tew et al.\cite{Tew2007} Repeating this latter procedure here for ACVnZ basis sets, we obtained $A_6$(\textrm{ACVnZ})=1.279 using the Tew et al.\cite{Tew2007} data, and 1.267 from our own CCSD(F12*)/REF-\{g,h\} calculations extrapolated $L^{-7}$ (as per Kutzelnigg and Morgan\cite{Kutzelnigg1992a}) from the `reference' basis sets of Hill et al.\cite{Hill2009} We note that a 0.05 discrepancy in $A_6$ for the W4-08 dataset will cause an average difference of just 0.03 kcal/mol, so it can safely be stated that the extrapolation coefficient is reasonably stable.

One minor detail must be mentioned in passing: the subvalence contribution listed in the W4-17 paper\cite{jmlm273} for OCS is erroneous owing to an atom transposition in the geometry input. The present calculations do not suffer from this issue. (A previously detected problem with the (Q) for FOOF was already reported and corrected in  Ref.\cite{jmlm330})

\subsection{Effect of subvalence post-CCSD(T) on thermochemistry}

Full data for the W4-08 dataset are available in the Supporting Information. An illustrative sample of the larger values is given in Table~\ref{tab:cv_exact}.

\begin{table*}[tp]
\caption{Selected species from the W4-11 dataset for illustration of post-CCSD(T) valence and core-valence correlation contributions to the TAE (kcal/mol). CCSD(T)/pwCVQZ reference geometries were used throughout.\label{tab:cv_exact}}
{\footnotesize 
\begin{tabular}{lllrrrrrrr}
\hline\hline
\textbf{pwCVTZ} &
  \textbf{Name of} &
  \textbf{} & 
  \textbf{pwCVTZ} &
  \textbf{pwCVTZ} &
  \textbf{lin. regression} &
  \textbf{pwCVDZ} &
  \textbf{pwCVTZ} &
  \textbf{pwCVTZ} & \textbf{cc-pVDZ}\\
\textbf{Val (Q)} &
  \textbf{the species} &
  \textbf{} &
  \textbf{CV T$_3$-(T)} &
  \textbf{CV (Q)} &
  \textbf{from pwCVDZ} &
  \textbf{CV (Q)} &
  \textbf{Val (Q)$_\Lambda$ - (Q)} &
  \textbf{CV (Q)$_\Lambda$ - (Q)} & \textbf{Val Q(5)$_\Lambda$ - (Q)$_\Lambda$}\\
  \hline
5.130 &
  tetrasulfur &
  S$_4$ (C$_{2V}$) &
  \cellcolor[HTML]{FAA5A7}0.123 &
  \cellcolor[HTML]{63BE7B}0.379 &
  {\color[HTML]{A02B93} 0.381} &
  \cellcolor[HTML]{5A8AC6}0.292 &
  \cellcolor[HTML]{93D1A5}-0.602 & &
  {\color[HTML]{0093AF} 0.256} \\ 

3.248 &
  dicarbon  &
  C$_2$ &
  \cellcolor[HTML]{F8696B}0.224 &
  \cellcolor[HTML]{D8EEE0}0.092 &
  {\color[HTML]{A02B93} 0.096} &
  \cellcolor[HTML]{DEE7F5}0.074 &
  \cellcolor[HTML]{9DD5AD}-0.574 &
  \cellcolor[HTML]{B8CCE7}-0.021 & 
  {\color[HTML]{0093AF} -0.240} \\ 
3.123 &
  boron nitride &
  BN ($^1\Sigma^+$) &
  \cellcolor[HTML]{FA9092}0.161 &
  \cellcolor[HTML]{CBE8D4}0.134 &
  {\color[HTML]{A02B93} 0.104} &
  \cellcolor[HTML]{DDE7F5}0.079 &
  \cellcolor[HTML]{6FC386}-1.001 &
  \cellcolor[HTML]{6390C9}-0.054 &
  {\color[HTML]{0093AF} -0.152} \\ 
  3.653 &
  dioxygen fluoride  &
  FO$_2$ &
  \cellcolor[HTML]{F9787A}0.200 &
  \cellcolor[HTML]{DBEFE3}0.084 &
  {\color[HTML]{A02B93} 0.080} &
  \cellcolor[HTML]{E5ECF7}0.062 &
  \cellcolor[HTML]{63BE7B}-0.932 &
  \cellcolor[HTML]{B9CCE9}-0.022 &
  {\color[HTML]{0093AF} 0.083} \\

2.508 &
  trisulfur  &
  S$_3$ &
  \cellcolor[HTML]{FBD5D8}0.048 &
  \cellcolor[HTML]{A1D8B1}0.226 &
  {\color[HTML]{A02B93} 0.227} &
  \cellcolor[HTML]{A1BCDF}0.174 &
  \cellcolor[HTML]{D3EBDB}-0.228 &
   \cellcolor[HTML]{B8CCE7}-0.017 &
   {\color[HTML]{0093AF} 0.088}\\
2.682 &
  tetraphosphorus &
  P$_4$ &
  \cellcolor[HTML]{FBD4D7}0.051 &
  \cellcolor[HTML]{B1DEBF}0.187 &
  {\color[HTML]{A02B93} 0.158} &
  \cellcolor[HTML]{C1D3EB}0.121 &
  \cellcolor[HTML]{F6F9FA}-0.014 & &

   {\color[HTML]{0093AF} -0.066}\\
4.361 &
  ozone &
  O$_3$ &
  \cellcolor[HTML]{FAA1A3}0.133 &
  \cellcolor[HTML]{E0F1E7}0.071 &
  {\color[HTML]{A02B93} 0.072} &
  \cellcolor[HTML]{E8EEF8}0.055 &
  \cellcolor[HTML]{A6D9B5}-0.505 &
  \cellcolor[HTML]{DEE7F4}-0.009 &
  {\color[HTML]{0093AF} 0.009}\\
2.164 &
  disulfur oxide &
  S$_2$O &
  \cellcolor[HTML]{FBCACD}0.066 &
  \cellcolor[HTML]{C9E7D3}0.128 &
  {\color[HTML]{A02B93} 0.146} &
  \cellcolor[HTML]{C6D6EC}0.112 &
  \cellcolor[HTML]{DAEEE2}-0.184 &
  \cellcolor[HTML]{D6E1F1}-0.011 &
 {\color[HTML]{0093AF} -0.032}\\

3.631 &
  dioxygen   difluoride &
  FOOF &
  \cellcolor[HTML]{FAABAD}0.117 &
  \cellcolor[HTML]{E6F3EC}0.057 &
  {\color[HTML]{A02B93} 0.059} &
  \cellcolor[HTML]{EEF2FA}0.045 &
  \cellcolor[HTML]{B9E0C5}-0.395 & 
  \cellcolor[HTML]{E7EDF7}-0.006 &
  {\color[HTML]{0093AF} 0.000} \\
1.887 &
  sulfur trioxide &
  SO$_3$ &
  \cellcolor[HTML]{FAACAE}0.115 &
  \cellcolor[HTML]{E9F5EF}0.047 &
  {\color[HTML]{A02B93} 0.087} &
  \cellcolor[HTML]{E1E9F6}0.067 &
  \cellcolor[HTML]{ECF5F1}-0.076 & 
  \cellcolor[HTML]{E3EAF6}-0.007 &
  {\color[HTML]{0093AF} -0.147} \\
1.808 &
  carbon disulfide &
  CS$_2$ &
  \cellcolor[HTML]{FCEBED}0.015 &
  \cellcolor[HTML]{C2E5CD}0.146 &
  {\color[HTML]{A02B93} 0.123} &
  \cellcolor[HTML]{D1DEF0}0.094 &
  \cellcolor[HTML]{F0F7F4}-0.058 &
  \cellcolor[HTML]{DDE6F4}-0.009 &
  {\color[HTML]{0093AF} 0.009} \\
1.428 &
  diphosphorus & 
  P$_2$ &
  \cellcolor[HTML]{FCE2E4}0.028 &
  \cellcolor[HTML]{D1EBDA}0.109 &
  {\color[HTML]{A02B93} 0.101} &
  \cellcolor[HTML]{DBE5F4}0.077 &
  \cellcolor[HTML]{ECF5F1}-0.078 &
  \cellcolor[HTML]{F4F6FC}-0.002 &
  {\color[HTML]{0093AF} 0.072}\\
1.177 &
  diboron &
  B$_2$ ($^3\Sigma^-_g$) &
  \cellcolor[HTML]{FBD7DA}0.047 &
  \cellcolor[HTML]{D9EEE1}0.090 &
  {\color[HTML]{A02B93} 0.079} &
  \cellcolor[HTML]{E6EDF8}0.060 &
  \cellcolor[HTML]{CFEAD8}-0.261 &
  \cellcolor[HTML]{A8C1E1}-0.026&
  {\color[HTML]{0093AF} 0.341}\\
1.737 &
  sulfur dioxide &
  SO$_2$ &
  \cellcolor[HTML]{FBC0C3}0.082 &
  \cellcolor[HTML]{E8F4EE}0.051 &
  {\color[HTML]{A02B93} 0.081} &
  \cellcolor[HTML]{E4EBF7}0.062 &
  \cellcolor[HTML]{E5F3EB}-0.116 &
  \cellcolor[HTML]{E7EDF7}-0.006 &
  {\color[HTML]{0093AF} -0.119}\\

1.932 &
  chlorine dioxide &
  OClO &
  \cellcolor[HTML]{FBCBCE}0.065 &
  \cellcolor[HTML]{E5F3EB}0.058 &
  {\color[HTML]{A02B93} 0.080} &
  \cellcolor[HTML]{E5ECF7}0.061 &
  \cellcolor[HTML]{DCEFE4}-0.170 &
  \cellcolor[HTML]{E8EEF8}-0.006 &
  {\color[HTML]{0093AF} -0.087}\\
2.124 &
  nitrogen dioxide &
  NO$_2$ &
  \cellcolor[HTML]{FBC8CA}0.072 &
  \cellcolor[HTML]{E9F4EE}0.049 &
  {\color[HTML]{A02B93} 0.050} &
  \cellcolor[HTML]{F2F5FC}0.038 &
  \cellcolor[HTML]{DEEFE5}-0.167 &
  \cellcolor[HTML]{E4EBF6}-0.007 &
  {\color[HTML]{0093AF} -0.049}\\
2.148 &
  hydrogen azide &
  HN$_3$ &
  \cellcolor[HTML]{FBC3C5}0.079 &
  \cellcolor[HTML]{EDF6F2}0.039 &
  {\color[HTML]{A02B93} 0.043} &
  \cellcolor[HTML]{F6F8FD}0.033 &
  \cellcolor[HTML]{E5F3EC}-0.120 &
  \cellcolor[HTML]{EDF1F9}-0.004&
  {\color[HTML]{0093AF} -0.051}\\
2.255 &
  nitrous oxide &
  N$_2$O &
  \cellcolor[HTML]{FBC6C9}0.074 &
  \cellcolor[HTML]{ECF6F1}0.042 &
  {\color[HTML]{A02B93} 0.045} &
  \cellcolor[HTML]{F5F7FD}0.034 &
  \cellcolor[HTML]{E2F1E8}-0.143 &
  \cellcolor[HTML]{EBF0F9}-0.005&
  {\color[HTML]{0093AF} -0.096} \\
1.457 &
  cyano radical &
  CN &
  \cellcolor[HTML]{FBC4C7}0.078 &
  \cellcolor[HTML]{EFF7F4}0.033 &
  {\color[HTML]{A02B93} 0.034} &
  \cellcolor[HTML]{FAFAFE}0.026 &
  \cellcolor[HTML]{C8E7D2}-0.306 &
  \cellcolor[HTML]{D3DFF0}-0.013&
  {\color[HTML]{0093AF} -0.060}\\
2.410 &
  cyanogen &
  NCCN &
  \cellcolor[HTML]{FBC3C6}0.078 &
  \cellcolor[HTML]{F0F7F4}0.032 &
  {\color[HTML]{A02B93} 0.038} &
  \cellcolor[HTML]{F8F9FE}0.029 &
  \cellcolor[HTML]{ECF5F1}-0.082 &
  \cellcolor[HTML]{F6F7FC}-0.002&
  {\color[HTML]{0093AF} -0.012}\\
1.415 &
  carbon oxide   sulfide &
  OCS &
  \cellcolor[HTML]{FCDCDF}0.039 &
  \cellcolor[HTML]{E0F1E7}0.072 &
  {\color[HTML]{A02B93} 0.063} &
  \cellcolor[HTML]{ECF1FA}0.049 &
  \cellcolor[HTML]{F0F7F5}-0.054 &
  \cellcolor[HTML]{E8EEF8}-0.006&
  {\color[HTML]{0093AF} -0.053} \\

1.924 &
  fulminic acid &
  HCNO &
  \cellcolor[HTML]{FBD0D2}0.058 &
  \cellcolor[HTML]{EDF6F2}0.038 &
  {\color[HTML]{A02B93} 0.038} &
  \cellcolor[HTML]{F8F9FE}0.029 &
  \cellcolor[HTML]{DFF0E6} -0.160 &
  \cellcolor[HTML]{E9EFF8}-0.006 & 
  {\color[HTML]{0093AF} -0.069}\\
1.620 &
  C-nitrous acid &
  C-HONO &
  \cellcolor[HTML]{FBCFD2}0.060 &
  \cellcolor[HTML]{EFF7F4}0.032 &
  {\color[HTML]{A02B93} 0.035} &
  \cellcolor[HTML]{F9FAFE}0.027 &
  \cellcolor[HTML]{E7F3ED}-0.099 &
  \cellcolor[HTML]{F0F3FA}-0.003&
  {\color[HTML]{0093AF} -0.058}\\
1.652 &
  T- nitrous acid &
  T-HONO &
  \cellcolor[HTML]{FBCFD2}0.059 &
  \cellcolor[HTML]{F0F7F4}0.033 &
  {\color[HTML]{A02B93} 0.036} &
  \cellcolor[HTML]{F9FAFE}0.028 &
  \cellcolor[HTML]{E9F4EE}-0.112 &
  \cellcolor[HTML]{F2F5FB}-0.003&
  {\color[HTML]{0093AF} -0.054}\\

1.189 &
  isocyanic acid &
  HNCO &
  \cellcolor[HTML]{FBD5D7}0.051 &
  \cellcolor[HTML]{EFF7F4}0.033 &
  {\color[HTML]{A02B93} 0.033} &
  \cellcolor[HTML]{FAFBFF}0.025 &
  \cellcolor[HTML]{F1F7F6}-0.046 &
  \cellcolor[HTML]{EBF0F9}-0.005 &
  {\color[HTML]{0093AF} -0.070}\\

0.547 &
  tetrafluorosilane &
  SiF$_4$ &
  \cellcolor[HTML]{FBBEC1}0.086 &
  \cellcolor[HTML]{FCFCFF}-0.003 &
  {\color[HTML]{A02B93} 0.030} &
  \cellcolor[HTML]{FBFCFF}0.023 &
  \cellcolor[HTML]{F8FAFC}-0.004 & \\

1.002 &
  silicon monoxide &
  SiO &
  \cellcolor[HTML]{FBD2D5}0.054 &
  \cellcolor[HTML]{F1F8F5}0.028 &
  {\color[HTML]{A02B93} 0.052} &
  \cellcolor[HTML]{F1F4FB}0.040 &
  \cellcolor[HTML]{E3F2E9}-0.133 &
  \cellcolor[HTML]{E7EDF7}-0.006 &
  {\color[HTML]{0093AF} -0.091} \\

1.223 &
  dichlorine   monoxide &
  Cl$_2$O &
  \cellcolor[HTML]{FCE2E5}0.029 &
  \cellcolor[HTML]{E8F4EE}0.052 &
  {\color[HTML]{A02B93} 0.054} &
  \cellcolor[HTML]{F1F4FB}0.041 &
  \cellcolor[HTML]{F2F7F6}-0.045 &
  \cellcolor[HTML]{FAFAFE}-0.001 &
   {\color[HTML]{0093AF} -0.069} \\
0.983 &
  carbon dichloride &
  CCl$_2$ &
  \cellcolor[HTML]{FCFBFE}-0.011 &
  \cellcolor[HTML]{D8EEE0}0.090 &
  {\color[HTML]{A02B93} 0.066} &
  \cellcolor[HTML]{EBF0F9}0.051 &
  \cellcolor[HTML]{EFF6F4}-0.061 &
  \cellcolor[HTML]{E8EEF8}-0.006 &
  {\color[HTML]{0093AF} -0.006} \\

1.174 &
  carbon dioxide &
  CO$_2$ &
  \cellcolor[HTML]{FCD8DB}0.044 &
  \cellcolor[HTML]{EFF7F4}0.033 &
  {\color[HTML]{A02B93} 0.032} &
  \cellcolor[HTML]{FAFBFF}0.025 &
  \cellcolor[HTML]{F1F7F5}-0.049 &
  \cellcolor[HTML]{EAEFF8}-0.005&
  {\color[HTML]{0093AF} -0.102} \\

1.003 &
  carbon sulfide &
  CS &
  \cellcolor[HTML]{FCF9FC}-0.007 &
  \cellcolor[HTML]{DBEFE2}0.084 &
  {\color[HTML]{A02B93} 0.065} &
  \cellcolor[HTML]{ECF1FA}0.050 &
  \cellcolor[HTML]{EFF6F4}-0.061 &
  \cellcolor[HTML]{E4EBF6}-0.007 &
  {\color[HTML]{0093AF} 0.022}\\

1.404 &
  glyoxal &
  C$_2$H$_2$O$_2$ &
  \cellcolor[HTML]{FBD6D8}0.049 &
  \cellcolor[HTML]{F1F8F6}0.028 &
  {\color[HTML]{A02B93} 0.029} &
  \cellcolor[HTML]{FCFCFF}0.023 &
  \cellcolor[HTML]{F0F7F5}-0.054 &
  \cellcolor[HTML]{F2F5FB}-0.003 \\
1.469 &
  difluorine monoxide &
  F$_2$O &
  \cellcolor[HTML]{FBCDD0}0.050 &
  \cellcolor[HTML]{FAFCFE}0.020 &
  {\color[HTML]{A02B93} 0.026} &
  \cellcolor[HTML]{F9FAFE}0.020 &
  \cellcolor[HTML]{F6F9FA}-0.089 &
  \cellcolor[HTML]{F0F3FA}-0.002&
  {\color[HTML]{0093AF} -0.027}\\
1.215 &
  isofulminic acid &
  HONC &
  \cellcolor[HTML]{FCD9DC}0.043 &
  \cellcolor[HTML]{F2F8F6}0.027 &
  {\color[HTML]{A02B93} 0.028} &
  \cellcolor[HTML]{FCFCFF}0.021 &
  \cellcolor[HTML]{F3F8F7}-0.035 &
  \cellcolor[HTML]{F6F8FD}-0.002 &
  {\color[HTML]{0093AF} -0.028}\\

1.342 &
  cyanogen chloride &
  ClCN &
  \cellcolor[HTML]{FCE0E3}0.032 &
  \cellcolor[HTML]{EEF7F3}0.035 &
  {\color[HTML]{A02B93} 0.037} &
  \cellcolor[HTML]{F8FAFE}0.028 &
  \cellcolor[HTML]{F4F9F8}-0.030 &
  \cellcolor[HTML]{FBFBFE}0.000&
  {\color[HTML]{0093AF} -0.031}\\

0.437 &
  aluminium   trifluoride &
  AlF$_3$ &
  \cellcolor[HTML]{FBCDD0}0.062 &
  \cellcolor[HTML]{FAFCFE}0.005 &
  {\color[HTML]{A02B93} 0.035} &
  \cellcolor[HTML]{F9FAFE}0.027 &
  \cellcolor[HTML]{F6F9FA}-0.018 &
  \cellcolor[HTML]{F0F3FA}-0.003&
  {\color[HTML]{0093AF} -0.068}\\

0.817 &
  disulfur &
  S$_2$ &
  \cellcolor[HTML]{FCFCFF}-0.014 &
  \cellcolor[HTML]{E1F1E7}0.070 &
  {\color[HTML]{A02B93} 0.063} &
  \cellcolor[HTML]{ECF1FA}0.048 &
  \cellcolor[HTML]{F4F9F8}-0.029 &
  \cellcolor[HTML]{FBFBFE}0.000&
  {\color[HTML]{0093AF} 0.025}\\

0.533 &
  aluminium   trichloride &
  AlCl$_3$ &
  \cellcolor[HTML]{FCF5F8}-0.003 &
  \cellcolor[HTML]{E5F3EB}0.058 &
  {\color[HTML]{A02B93} 0.055} &
  \cellcolor[HTML]{F0F4FB}0.042 &
  \cellcolor[HTML]{FCFCFF}0.016 & &
   {\color[HTML]{0093AF} -0.006}\\

0.875 &
  sulfur monoxide &
  SO &
  \cellcolor[HTML]{FCE3E6}0.027 &
  \cellcolor[HTML]{F3F8F7}0.024 &
  {\color[HTML]{A02B93} 0.036} &
  \cellcolor[HTML]{F9FAFE}0.027 &
  \cellcolor[HTML]{F2F8F6}-0.042 &
  \cellcolor[HTML]{F8F9FD}-0.001 &
  {\color[HTML]{0093AF} -0.024}\\
0.580 &
  hydrogen disulfide   radical &
  SSH &
  \cellcolor[HTML]{FCF4F7}0.000 &
  \cellcolor[HTML]{E8F4EE}0.049 &
  {\color[HTML]{A02B93} 0.042} &
  \cellcolor[HTML]{F6F8FD}0.032 &
  \cellcolor[HTML]{F4F8F8}-0.030 &
  \cellcolor[HTML]{F6F8FD}-0.001 &
  {\color[HTML]{0093AF} 0.008}\\
0.650 &
  chlorine monoxide &
  ClO &
  \cellcolor[HTML]{FCEDF0}0.012 &
  \cellcolor[HTML]{EEF7F3}0.035 &
  {\color[HTML]{A02B93} 0.031} &
  \cellcolor[HTML]{FBFBFF}0.024 &
  \cellcolor[HTML]{ECF5F1}-0.081 &
  \cellcolor[HTML]{EBF0F9}-0.005 &
  {\color[HTML]{0093AF} 0.007} \\

0.910 &
difluoride &
F$_2$ &
\cellcolor[HTML]{FCFBFE}0.036 &
\cellcolor[HTML]{EDF6F2}0.007 &
{\color[HTML]{A02B93} 0.012} &
\cellcolor[HTML]{FAFBFF}0.010 &
\cellcolor[HTML]{F8FAFC}-0.048 &
\cellcolor[HTML]{FCF9FC}0.000 &
{\color[HTML]{0093AF} -0.021}\\

0.147 &
 disilane   &
Si$_2$H$_6$ &
\cellcolor[HTML]{FBC3C5}0.081 &
\cellcolor[HTML]{F8FAFC}-0.052 &
{\color[HTML]{A02B93} -0.032} &
\cellcolor[HTML]{FAFBFF}-0.025 &
\cellcolor[HTML]{F8FAFC}0.001 & 
\cellcolor[HTML]{FCF9FC}0.007 &
{\color[HTML]{0093AF} 0.003}\\

0.447 &
  dichlorine &
  Cl$_2$ &
  \cellcolor[HTML]{FCFBFE}0.012 &
  \cellcolor[HTML]{EDF6F2}0.039 &
  {\color[HTML]{A02B93} 0.034} &
  \cellcolor[HTML]{FAFBFF}0.026 &
  \cellcolor[HTML]{F8FAFC}-0.005 &
  \cellcolor[HTML]{FCF9FC}0.001 &
  {\color[HTML]{0093AF} 0.006}\\
  \hline\hline
\end{tabular}

}

\end{table*}

(a) For valence correlation, it has already been established\cite{jmlm330} that CCSDT(Q)$_\Lambda$ recovers the lion’s share of the post-CCSDT(Q) correlation effects. As expected, the valence (Q)$_\Lambda$ - (Q) difference is largest for species with significant static correlation: -0.57 kcal/mol for \ce{C2}, -1.00 for \ce{BN}, -0.26 for \ce{B2}, -0.50 for ozone, -0.60 for \ce{S4}, -0.23 for \ce{S3}, -0.39 for \ce{FOOF}. Additional species with lesser degrees of static correlation include \ce{CN} radical (-0.31 kcal/mol), \{-0.13, -0.14, -0.17,  -0.17, -0.17,  -0.18 \} kcal/mol for \{\ce{SiO}, \ce{N2O}, \ce{NO2}, \ce{BN} ($a^3\Pi$), \ce{OClO}, \ce{S2O} \}.

(b) That said, we evaluated the differential effect of core-valence correlation on the (Q)$_\Lambda$ - (Q) difference for the W4-08 subset. For the most part it is negligible (0.01 kcal/mol or less), and just for a handful of species such as \ce{B2} (-0.026), BN (-0.054), \ce{C2} (-0.021 kcal/mol), and CN radical (-0.013 kcal/mol) somewhat larger values are seen. The former three species of course have strong static correlation, while CN’s UHF reference function is severely spin-contaminated.

(c) Let us now turn to the $\Delta\textrm{CV}$(Q) core-valence contributions. In first-row species, these are significant only for species with strong static correlation --- such as the usual suspects \{\ce{C2}, \ce{BN}, \ce{B2}, \ce{O3}\} at \{0.09, 0.13, 0.09, 0.07\} kcal/mol --- plus lesser, yet still nontrivial, contributions for FOOF, \ce{N2O}, \ce{NO2}, and the like. These are amounts comparable to the RMS uncertainties of W4 theory\cite{jmlm200} and especially W4-F12 theory,\cite{jmlm269} but are still on the edge of tolerable.

In contrast, in the second row, large $\Delta\textrm{CV}$(Q) contributions are also seen for species that are fundamentally single-reference, such as 0.19 kcal/mol for \ce{P4}  --- 0.11 kcal/mol for \ce{P2} could be partially attributed to static correlation, as can (less plausibly) 0.08 kcal/mol for \ce{CS} and 0.15 kcal/mol for \ce{CS2}. But the largest effects are seen in species that are both multireference and second-row, reaching a whopping 0.38 kcal/mol for \ce{S4} and 0.23 kcal/mol for \ce{S3}, contrasting with just 0.07 kcal/mol for \ce{S2}.

Contrariwise, in species like \ce{SiF4}, and \ce{AlF3} --- which have central second-row atoms with smallish core-valence gaps, but no adjacent pairs of second-row atoms --- $\Delta\textrm{CV}$(Q) is negligible at -0.003 and +0.005 kcal/mol, respectively. But in  \ce{AlCl3}, which does have such pairs, (Q) is somewhat more significant at 0.06 kcal/mol. Likewise, it reaches 0.09 kcal/mol in \ce{CCl2}.

At the other extreme from species like \ce{S4}, in the silanes (which exhibit essentially pure dynamical correlation) there are small negative (antibonding) contributions to the TAE, as the molecule is in fact less `multireference’ than silicon atom.

(d) There is much less of a difference between first- and second-row for the core-valence contribution of higher-order triple excitations, $T_3 - (T)$. We find the expected large contributions for species with strong static correlation — for instance, 0.22 kcal/mol for \ce{C2} and 0.13 kcal/mol for \ce{O3}, both of which already commented on in the original W4 papers for the old W4-17 geometries.\cite{jmlm200,jmlm205}  But for \ce{S4} we find 0.12 kcal/mol, not dissimilar from 0.12 kcal/mol for FOOF. On the other hand, we see 0.08 kcal/mol for disilane, which admittedly is partially canceled by the negative quadruples contribution.
 In addition, now radical species with a degree of UHF spin contamination make an entry, such as 0.08 kcal/mol for CN radical.

In the original W4 paper,\cite{jmlm200} an average `ACES — MOLPRO difference correction’ was applied to the CCSD(T) energy, which in fact can be regarded as a primitive estimate for core-valence $T_3 - (T)$. 
In the present work, we sidestepped the issue by using UHF references throughout, for which orbitals are canonical to begin with and the said correction term thus identically zero. 

(e) The sum of $T_3-(T)$ and $(T_4)$ correlation, i.e., the CCSDT(Q) - CCSD(T) difference, reflects a high degree of synergy between the two components. It reaches a maximum of 0.50 kcal/mol for \ce{S4}, with fairly hefty values of \{0.27, 0.24\} also seen for \{\ce{S3}, \ce{P4}\} but once can still see \{0.13, 0.16\} kcal/mol for \{\ce{SO2}, \ce{SO3}\}. Among 2nd-row diatomics, \ce{P2} stands out (see also Persson et al.\cite{JoakimPersson1997} and Ref.\cite{jmlm207}).
Large contributions are also seen for a few first-row systems with strong static correlation, such as \{\ce{BN}, \ce{C2}, \ce{O3}, \ce{FOOF}\} at \{0.30, 0.32 ,0.20, 0.17\} kcal/mol, although 0.12 kcal/mol for \ce{NO2} and \ce{N2O}, as well as 0.14 kcal/mol for \ce{B2} and 0.11 kcal/mol for \ce{CN} are also noteworthy. 

(f) The combined effects of geometry shift and core-valence post-CCSD(T) can reach 0.13, 0.19, and 0.19  kcal/mol, respectively, for CN, \ce{O3}, and \ce{P2}; the largest contributions reach 0.28 kcal/mol for \ce{SO3},
0.33 kcal/mol for \ce{S3}, 
0.38 kcal/mol for \ce{P4}, 
and a whopping 0.55 kcal/mol for \ce{S4}.

For the purpose of any next-generation successor to `W4 theory', the bottom line is this: simply evaluating core-valence contributions at the CCSD(T) level at a valence-optimized geometry may be adequate for most first-row systems, but clearly `has been weighed in the balance and found wanting' for second-row systems.

A final remark: clearly, subvalence CCSDT(Q)/cc-pwCVTZ calculations would become arduous for species with many second-row atoms such as \ce{C2Cl6}, and even for smaller species if they lack symmetry. Is there a more economical alternative, such as CCSDT(Q)/cc-pwCVDZ? As can be seen in Table~\ref{tab:cv_exact}, this small basis set underestimates the $\Delta$CV(Q) seriously, but fairly systematically: scaling by 1.30 leads to an RMSD of just 0.009 kcal/mol with the more rigorous values. This may hence be a practical option for larger systems.

And while we are on the subject: are values adequately converged with the basis set for pwCVTZ? While the computational cost for subvalence CCSDT(Q)/cc-pwCVQZ is prohibitive for species like \ce{S4}, we were able, at great cost, to obtain data for a subset of species (see ESI). The most expensive calculation in the batch, for \ce{SO3}, took 1 month of wall time on an Intel Ice Lake node with 52 cores, 768GB RAM, and 6TB SSD. The differential core-valence (Q) contribution to TAE does change somewhat (5-10\%)  between cc-pwCVTZ and cc-pwCVQZ, but in absolute numbers this change amounts to less than 0.01 kcal/mol.

\subsection{Revisiting valence post-CCSD(T) contributions}

At first, we considered recycling CCSDT(Q)/cc-pwCVTZ from the core-valence calculation and adding CCSDT(Q)/cc-pwCVTZ for a CCSDT(Q)/cc-pwCV\{T,Q\}Z calculation. However, the latter proved too taxing for several molecules; hence, we explored alternatives using valence correlation consistent basis sets.

CCSDT(Q)/cc-pV(\{T,Q\}+d)Z turned out to yield nearly identical results. While CCSDT(Q)/cc-pV(T+d)Z differed significantly from CCSDT(Q)/cc-pVTZ for some second-row molecules (notably 0.10 kcal/mol for \ce{P4}), upon extrapolation  the differences and cc-pV\{T,Q\}Z and cc-pV(\{T,Q\}+d)Z basically disappear, as already noted by Karton\cite{Karton2022b}. We also carried out cc-pV5Z calculations for a large subset, and found the cc-pV\{Q,5\}Z and cc-pV\{T,Q\}Z post-CCSD(T) corrections to agree to about 0.01 kcal/mol RMS. Note that Schwenke coefficients for the extrapolation were taken from the earlier work of Karton\cite{Karton2020}.

\subsection{Post-CCSDT(Q)$_\Lambda$ contributions}

While it has already been established\cite{jmlm326} that CCSDT(Q)$_\Lambda$ is superior to CCSDT(Q) and indeed CCSDTQ, some residual higher-order contributions remain. From the ESI of Ref.\cite{jmlm330}, we find an RMS CCSDTQ5(6)$_\Lambda$-CCSDTQ(5)$_\Lambda$ difference of just 0.012 kcal/mol with the un-polarized cc-pVDZ(p,s) basis set, indicating that CCSDTQ(5)$_\Lambda$ is adequately close to the FCI limit for our purposes.

The CCSDTQ(5)$_\Lambda$-CCSDT(Q)$_\Lambda$ difference, however, is still of some significance, reaching 0.097 kcal/mol RMS for the cc-pVDZ(p,s) basis set and just 0.069 kcal/mol for cc-pVDZ(d,s). The difference between the RMS values for unpolarized and polarized basis sets is almost entirely due to the \ce{C2} molecule, for which the unpolarized basis set is simply too anemic.

The effect of different reference geometries on this quantity is clearly negligible, hence the RMS CCSDTQ(5)$_\Lambda$-CCSDT(Q)$_\Lambda$/cc-pVDZ contribution at the new geometry is almost identical at 0.067 kcal/mol.

\subsection{While we are at it: scalar relativistics and DBOC reconsidered}

In the original W4 theory, scalar relativistic effects were treated by 2nd-order Douglas-Kroll-Hess (DKH2) at the valence CCSD(T) level using the aug-cc-pV(Q+d)Z basis set and its relativistic recontraction. 

Here, we have considered the aug-cc-pCV$n$Z (n=T,Q,5) basis sets using the X2C (exact two-component\cite{ReiherX2C}) treatment as implemented in MOLPRO. 

As we previously found in Ref.\cite{jmlm283}, we find no significant differences between DKH2 and X2C for the present systems, not even for cases like \ce{AlCl3}.

There are slight differences (ca. 0.01 kcal/mol) between aug-cc-pCVQZ and aug-cc-pCV5Z for some systems, but by and large, the scalar relativistic components from the W4-17 paper\cite{jmlm273} are the same as what we obtained presently.

What happens if we permit subvalence correlation? $\Delta \textrm{CV}\Delta \textrm{REL}$, the differential subvalence-relativistic contribution to TAE, is found to be insignificant for the first row, but for some second-row species it gets to be less trivial: \{-0.07,-0.05,-0.07\} kcal/mol for \{\ce{AlCl3},\ce{AlF3},\ce{Si2H6}\} are standouts, but one also sees -0.03 kcal/mol for \ce{AlH3}, \ce{S4}, and \ce{SO3}, and -0.04 kcal/mol for \ce{SiH4}. 

While we were at it, we considered also the DBOC, which for obvious reasons is most important for species with many hydrogens.

It is fairly well known (e.g., Gauss and coworkers\cite{Gauss2006DBOC,jmlm203}) that electron correlation will reduce DBOCs by about half. Indeed, we observe here that correlation reduces DBOC contributions to TAEs across the board, and indeed even pushes them in negative (antibonding) territory for some species.

As can be seen in the ESI, basis set sensitivity is rather modest, with aug-cc-pCVTZ being clearly adequate. Adding diffuse functions was found to affect DBOCs only insignificantly, while the same is largely true for including subvalence correlation.

Thorpe and Stanton already noted\cite{StantonDBOC} that for some species, like \ce{NO} and \ce{NO2}, DBOCs calculated at any level will be highly suspect because the equilibrium geometries are near Hartree-Fock instabilities. Specifically for \ce{NO2}, we found an absurdly large DBOC at both the W4-17 and HEAT reference geometries; a potential surface scan revealed that the DBOC exhibits a `pole' near these geometries. When displacing the angle by a few degree, the DBOC levels off at -0.05 kcal/mol, but even that value should be taken with a grain of salt. Ultimately, we elected to suppress the DBOC for \ce{NO2} altogether.

\subsection{Rotational zero-point energy}

A handful of species are nontrivially affected by a phenomenon first pointed out for OH radical by Ruscic and coworkers\cite{Ruscic2002_OH_H2O_BDE} and explained in more detail by Stanton and coworkers\cite{HEAT3}, by Ruscic in Ref.\cite{Klippenstein2017_ATcT_CoreCombustion}, and on pp. 16--17 of Ruscic and Bross's thermochemistry review\cite{Ruscic2019}.

Focusing here on diatomics for the moment, 
the lowest rotationless energy of a molecule is not necessarily identical to the lowest \emph{allowed} rotational level, as the rotational ground state may be forbidden by spin and/or spatial symmetry. 

The most significant examples here are for several diatomics with spin-orbit splitting. Particularly for open-shell species in degenerate states, coupling between rotation and spin-orbit splitting leads to the lowest allowed rotational energy level (LAREL) differing nontrivially (by thermochemical standards) from the lowest rotation-free spin-orbit level. The difference is referred to as `rotational zero-point energy'.\footnote{Stanton\cite{HEAT3} used this term in the present context, but it was previously used by Adams and Smith\cite{AdamsSmith1981} when referring to the very clear note by Carney and Porter\cite{Carney1977LowestH3} on the lowest vibrational levels of \ce{H3+}.} The LAREL  can be calculated by the Hill-Van Vleck equation (lemma V.28 in Herzberg\cite{HerzbergVol1}, see also Hougen\cite{HougenNBS115}); using rotational constants and spin-orbit splittings from Huber and Herzberg\cite{HuberHerzberg1979} and comparing with the rotation-free level (which is also recovered from the Hill-Van Vleck equation in the low-rotational constant limit), we obtain adjustments of the dissociation energy by -0.09 kcal/mol for OH, -0.05 for SH, -0.04 kcal/mol for CH, -0.02 for SiH, and less than 0.01 kcal/mol for remaining species. The results for OH and CH echo Refs.\cite{Ruscic2002_OH_H2O_BDE,Thorpe2023}

\subsection{A first attempt at a W5 protocol; comparison with earlier results}

We tentatively propose two preliminary `Weizmann-5' protocols, namely, W5prelim1 and W5prelim2:
\begin{itemize}
  \setlength{\itemsep}{0pt}
  \setlength{\parskip}{0pt}
  \setlength{\topsep}{0pt}
  \setlength{\partopsep}{0pt}
  \setlength{\parsep}{0pt}

    \item in all steps, unless explicitly indicated otherwise, all electrons correlated except for the (1s) `deep cores' in Al--Cl;
    \item UHF references used unless explicitly indicated otherwise;
    \item geometry optimized at CCSD(T)/cc-pwCVQZ level; for open-shell systems ROCCSD(T) rather than UCCSD(T);
    \item CCSD(T)/aug-cc-pCV\{5,6\}Z calculations with Schwenke-style two-point extrapolation. If subvalence CCSD(T)/aug-cc-pCV6Z impossible, then CCSD(T)/aug-cc-pCV\{5,6\}Z valence only and CCSD(T)/aug-cc-pCV\{Q,5\}Z subvalence;
    \item for post-CCSD(T) basis set extension, valence CCSDT(Q)/pV(n+d)Z (n=Q) and CCSDT(Q)/pV(n+d)Z (n=T) with two-point extrapolations following Karton\cite{Karton2020}. (cc-pV(n+d)Z (n= T, Q) were treated like cc-pVnZ);
    \item subvalence post-CCSD(T) from valence and subvalence CCSDT(Q)/cc-pwCVTZ;
    \item for post-CCSDT(Q) corrections, valence CCSDT(Q)$_\Lambda$/cc-pV(T+d)Z - CCSDT(Q)/cc-pV(T+d)Z;
    \item for W5prelim2, add furthermore CCSDTQ(5)$_\Lambda$-CCSDT(Q)$_\Lambda$/cc-pVDZ;
    \item scalar relativistic X2C-CCSD(T)/aug-cc-pCV5Z valence for W5prelim1 and with subvalence correlation for W5prelim2;
    \item DBOC at the CCSD/aug-cc-pwCVTZ level, subvalence correlation omitted;
    \item add in rotational zero-point correction if needed.
\end{itemize}

\subsection{Comparison for W4-08 with earlier W4-17 results and ATcT}

\begin{table}[]

\caption{Computed total atomization energies at 0 K (kcal/mol) for the W4-08 subset at the W5prelim\{1,2\} level compared with the older W4-17 benchmark data and the latest ATcT data with associated uncertainties\label{tab:ATcT}.}
{
\footnotesize 
\begin{tabular}{lrrrrl}
\hline\hline
             & \textbf{W5prelim1} & \textbf{W5prelim2}                       & \textbf{W4-17}                    & \multicolumn{2}{c}{\textbf{ATcT version 1.220}}                       \\

             & through (Q)$_\Lambda$      & through (5)$_\Lambda$                            &                                   &           &             \\
         \hline
             & TAE$_0$& {\color[HTML]{C00000} TAE$_0$} & {\color[HTML]{E97132} TAE$_0$} & TAE$_0$  & uncertainty \\
             \hline

         RMSD$*$    & \textbf{0.080} & \textbf{0.066}                       & \textbf{0.074}                    &  reference         &             \\
\hline
\ce{B2H6}                & 567.37         & {\color[HTML]{C00000} 567.37}        & {\color[HTML]{E97132} 567.53}     & 567.44    & 0.43        \\
\ce{BHF2}                & 398.73         & {\color[HTML]{C00000} 398.69}        & {\color[HTML]{E97132} 398.73}     &           &             \\
\ce{BF3}                 & 461.33         & {\color[HTML]{C00000} 461.26}        & {\color[HTML]{E97132} 461.32}     & 461.34    & 0.20        \\
\ce{C2H6}                & 666.17         & {\color[HTML]{C00000} 666.16}        & {\color[HTML]{E97132} 666.28}     & 666.20    & 0.03        \\
\ce{H2CN}                & 327.86         & {\color[HTML]{C00000} 327.87}        & {\color[HTML]{E97132} 327.95}     & 327.91    & 0.13        \\
\ce{NCCN}                & 491.63         & {\color[HTML]{C00000} 491.61}        & {\color[HTML]{E97132} 491.50}     & 491.34    & 0.10        \\
\ce{CH2NH2}              & 450.95         & {\color[HTML]{C00000} 450.94}        & {\color[HTML]{E97132} 451.02}     & 450.93    & 0.08        \\
\ce{CH3NH}               & 444.13         & {\color[HTML]{C00000} 444.13}        & {\color[HTML]{E97132} 444.22}     & 444.10    & 0.09        \\
\ce{CH3NH2}              & 542.13         & {\color[HTML]{C00000} 542.12}        & {\color[HTML]{E97132} 542.22}     & 542.21    & 0.05        \\
\ce{CF2}                 & 253.37         & {\color[HTML]{C00000} 253.32}        & {\color[HTML]{E97132} 253.26}     & 253.30    & 0.08        \\
\ce{N2H}                 & 216.40         & {\color[HTML]{C00000} 216.41}        & {\color[HTML]{E97132} 216.45}     & 216.32    & 0.11        \\
\ce{t-N2H2}              & 278.64         & {\color[HTML]{C00000} 278.64}        & {\color[HTML]{E97132} 278.68}     & 278.70    & 0.10        \\
\ce{N2H4}                & 404.67         & {\color[HTML]{C00000} 404.66}        & {\color[HTML]{E97132} 404.76}     & 404.77    & 0.11        \\

\ce{FOOF}                & 147.33         & {\color[HTML]{C00000} 147.33}        & {\color[HTML]{E97132}  146.89$^{c}$}     & 146.47    & 0.10        \\

\ce{AlF3}                & 423.83         & {\color[HTML]{C00000} 423.71}              & {\color[HTML]{E97132} 423.48}     &           &             \\
\ce{Si2H6}               & 503.08         & {\color[HTML]{C00000} 503.01}        & {\color[HTML]{E97132} 503.09}     &  503.07   & 0.41$^{e}$  \\
\ce{P4}                  & 286.61         & {\color[HTML]{C00000} 286.51}        & {\color[HTML]{E97132} 285.96}     &           &             \\
\ce{SO2}                 & 254.63         & {\color[HTML]{C00000} 254.50}        & {\color[HTML]{E97132} 254.42}     & 254.48    & 0.05        \\
\ce{SO3}                 & 336.45         & {\color[HTML]{C00000} 336.27}        & {\color[HTML]{E97132} 336.12}     & 336.31    & 0.07        \\
\ce{OCS}                 & 329.19         & {\color[HTML]{C00000} 329.12}        & {\color[HTML]{E97132} 

328.96$^f$}     & 328.67    & 0.08        \\
\ce{CS2}                 & 274.98         & {\color[HTML]{C00000} 274.96}        & {\color[HTML]{E97132} 274.67}     & 274.61    & 0.16        \\
\ce{S2O}                 & 203.81         & {\color[HTML]{C00000} 203.77}        & {\color[HTML]{E97132} 203.58}     & 203.43    & 0.12        \\
\ce{S3}                  & 164.23         & {\color[HTML]{C00000} 164.29}        & {\color[HTML]{E97132} 163.97}     & 164.01    & 0.20        \\
\ce{S4 (C_{2v})}         & 228.38         & {\color[HTML]{C00000} 228.61}        & {\color[HTML]{E97132} 228.15}     &           &             \\
\ce{CCl2}                & 172.83         & {\color[HTML]{C00000} 172.81}        & {\color[HTML]{E97132} 172.71}     & 172.26    & 0.15        \\
\ce{AlCl3}               & 305.98         & {\color[HTML]{C00000} 305.91}        & {\color[HTML]{E97132} 305.60}     &           &             \\
\ce{ClCN}                & 278.94         & {\color[HTML]{C00000} 278.90}        & {\color[HTML]{E97132} 278.79}     & 278.79    & 0.10        \\
\ce{OClO}                & 122.72         & {\color[HTML]{C00000} 122.63}        & {\color[HTML]{E97132} 122.32}     & 122.34    & 0.07        \\

\ce{Cl2O}                &  97.29         & {\color[HTML]{C00000}  97.26}        & {\color[HTML]{E97132} 96.94}      & 97.11     & 0.09        \\
\ce{BN (^3\Pi)}          & 103.54         & {\color[HTML]{C00000} 103.51}        & {\color[HTML]{E97132} 103.55}     & 103.57    & 0.27        \\
\ce{CF}                  & 130.40         & {\color[HTML]{C00000} 130.38}        & {\color[HTML]{E97132} 130.35}     & 130.37    & 0.03        \\
\ce{CH2C}                & 345.02         & {\color[HTML]{C00000} 345.05}        & {\color[HTML]{E97132} 345.10}     & 345.03    & 0.07        \\
\ce{CH2CH}               & 422.97         & {\color[HTML]{C00000} 422.97}        & {\color[HTML]{E97132} 423.06}     & 422.95    & 0.07        \\
\ce{C2H4}                & 532.02         & {\color[HTML]{C00000} 532.02}        & {\color[HTML]{E97132} 532.11}     & 532.03    & 0.03        \\
\ce{CH2NH}               & 414.35         & {\color[HTML]{C00000} 414.34}        & {\color[HTML]{E97132} 414.41}     & 414.36    &       0.08      \\
\ce{HCO}                 & 270.76         & {\color[HTML]{C00000} 270.72}        & {\color[HTML]{E97132} 270.66}     & 270.76    & 0.02        \\
\ce{H2CO}                & 357.49         & {\color[HTML]{C00000} 357.46}        & {\color[HTML]{E97132} 357.51}     & 357.48    & 0.02        \\
\ce{CO2}                 & 382.08         & {\color[HTML]{C00000} 381.97}        & {\color[HTML]{E97132} 381.94}     & 381.97    & 0.01        \\
\ce{HNO}                 & 196.86         & {\color[HTML]{C00000} 196.85}        & {\color[HTML]{E97132} 196.78}     & 196.83    & 0.03        \\
\ce{NO2}                 & 221.69$^d$     & {\color[HTML]{C00000} 221.63$^d$}        & {\color[HTML]{E97132} 221.61}     & 221.66    & 0.02        \\
\ce{N2O}                 & 263.53         & {\color[HTML]{C00000} 263.43}        & {\color[HTML]{E97132} 263.43}     & 263.40    & 0.03        \\
\ce{O3}                  & 142.41         & {\color[HTML]{C00000} 142.42}        & {\color[HTML]{E97132} 142.33}     & 142.48    & 0.01        \\
\ce{HOO}                 & 166.01         & {\color[HTML]{C00000} 166.00}        & {\color[HTML]{E97132} 165.97}     & 166.03    & 0.00        \\
\ce{HOOH}                & 252.13         & {\color[HTML]{C00000} 252.10}        & {\color[HTML]{E97132} 252.08}     & 252.19    & 0.01        \\
\ce{F2O}                 &  89.63         & {\color[HTML]{C00000}  89.61}         & {\color[HTML]{E97132} 89.42}      & 89.52     & 0.05        \\
\ce{HOCl}                & 156.86         & {\color[HTML]{C00000} 156.84}        & {\color[HTML]{E97132} 156.73}     & 156.87    & 0.01        \\
\ce{SSH}                 & 157.85         & {\color[HTML]{C00000} 157.84}        & {\color[HTML]{E97132} 157.68}     &           &             \\
\ce{B2 ($^3\Sigma^-_g$)} &  65.52         & {\color[HTML]{C00000}  65.86}         & {\color[HTML]{E97132} 65.85}      &           &             \\
\ce{BH}                  &  81.45         & {\color[HTML]{C00000}  81.46}         & {\color[HTML]{E97132} 81.49}      & 81.48     & 0.25        \\
\ce{BH3}                 & 264.71         & {\color[HTML]{C00000} 264.72}        & {\color[HTML]{E97132} 264.81}     & 264.79    & 0.22        \\
\ce{BN ($^1\Sigma^+$)}   & 102.86         & {\color[HTML]{C00000} 102.71}        & {\color[HTML]{E97132} 102.62}     &           &             \\
\ce{BF}                  & 179.99         & {\color[HTML]{C00000} 179.97}        & {\color[HTML]{E97132} 179.95}     &           &             \\
\ce{NH ($^3\Sigma^-$)}   &  78.30         & {\color[HTML]{C00000}  78.30}         & {\color[HTML]{E97132} 78.34}      & 78.36     & 0.04        \\
\ce{NH2}                 & 170.51         & {\color[HTML]{C00000} 170.51}        & {\color[HTML]{E97132} 170.58}     & 170.59    &      0.03       \\
\ce{HCN}                 & 303.18         & {\color[HTML]{C00000} 303.17}        & {\color[HTML]{E97132} 303.21}     & 303.14    &      0.02       \\
\ce{HOF}                 & 149.31         & {\color[HTML]{C00000} 149.29}        & {\color[HTML]{E97132} 149.32}     & 149.26    &       0.05      \\
\ce{AlH}                 &  70.86         & {\color[HTML]{C00000}  70.87}         & {\color[HTML]{E97132} 70.84}      &           &             \\
\ce{AlH3}                & 200.83         & {\color[HTML]{C00000} 200.80}        & {\color[HTML]{E97132} 200.91}     &           &             \\
\ce{AlF}                 & 161.83         & {\color[HTML]{C00000} 161.80}        & {\color[HTML]{E97132} 161.76}     &           &             \\
\ce{AlCl}                & 120.69         & {\color[HTML]{C00000} 120.69}        & {\color[HTML]{E97132} 120.63}     &           &             \\
\ce{SiH}                 &  70.52         & {\color[HTML]{C00000}  70.51}         & {\color[HTML]{E97132} 70.65}      &   70.54        &     0.21$^e$        \\
\ce{SiH4}                & 303.89         & {\color[HTML]{C00000} 303.85}        & {\color[HTML]{E97132} 304.16}     &   303.98        &  0.20$^e$           \\
\hline\hline
\end{tabular}
}
\end{table}

\addtocounter{table}{-1}
\begin{table}
\caption{(Continued)}

{\footnotesize
\begin{tabular}{lrrrrl}
\hline\hline
\ce{SiO}          &190.38          & {\color[HTML]{C00000} 190.27 }        & {\color[HTML]{E97132} 190.37}     &    190.14       &          0.19$^{e}$   \\
\ce{SiF}          &140.59          & {\color[HTML]{C00000} 140.55 }        & {\color[HTML]{E97132} 140.62}     &  140.52         & 0.20$^{e}$            \\
\ce{CS}           &169.68          & {\color[HTML]{C00000} 169.69 }        & {\color[HTML]{E97132} 169.59}     & 169.52    &      0.16       \\
\ce{H2}           &103.27          & {\color[HTML]{C00000} 103.27 }        & {\color[HTML]{E97132} 103.28}     & 103.27    &  0.00           \\
\ce{OH}           &101.69          & {\color[HTML]{C00000} 101.69 }        & {\color[HTML]{E97132} 101.76}     & 101.72    & 0.01        \\
\ce{HF}           &135.30          & {\color[HTML]{C00000} 135.28 }        & {\color[HTML]{E97132} 135.27}     & 135.27    &      0.01       \\
\ce{H2O}          &219.35          & {\color[HTML]{C00000} 219.34 }        & {\color[HTML]{E97132} 219.32}     & 219.36    &     0.01        \\
\ce{CH (^2$\Pi$)} & 79.94          & {\color[HTML]{C00000}  79.95 }         & {\color[HTML]{E97132} 79.99}      & 79.94     & 0.02        \\
\ce{CH2 (^3B1)}   &179.84          & {\color[HTML]{C00000} 179.84 }        & {\color[HTML]{E97132} 179.86}     & 179.82    &     0.02        \\
\ce{CH3}          &289.06          & {\color[HTML]{C00000} 289.06 }        & {\color[HTML]{E97132} 289.08}     & 289.10    &     0.01        \\
\ce{CH4}          &392.45          & {\color[HTML]{C00000} 392.45 }        & {\color[HTML]{E97132} 392.46}     & 392.46    &       0.01      \\
\ce{CCH}          &256.97          & {\color[HTML]{C00000} 257.02 }        & {\color[HTML]{E97132} 257.04}     & 256.94    & 0.04        \\
\ce{C2H2}         &388.65          & {\color[HTML]{C00000} 388.66 }        & {\color[HTML]{E97132} 388.70}     & 388.61    & 0.03        \\
\ce{NH3}          &276.53          & {\color[HTML]{C00000} 276.52 }        & {\color[HTML]{E97132} 276.53}     & 276.60    &     0.01        \\
\ce{C2}           &144.26          & {\color[HTML]{C00000} 144.01 }        & {\color[HTML]{E97132} 144.07}     & 144.07    & 0.02        \\
\ce{N2}           &225.02          & {\color[HTML]{C00000} 225.00 }        & {\color[HTML]{E97132} 225.00}     & 224.96    & 0.01        \\
\ce{CO}           &256.23          & {\color[HTML]{C00000} 256.20 }        & {\color[HTML]{E97132} 256.15}     & 256.22    & 0.01        \\
\ce{CN}           &178.24          & {\color[HTML]{C00000} 178.18 }        & {\color[HTML]{E97132} 178.18}     & 178.12    & 0.01        \\
\ce{NO}           &149.85          & {\color[HTML]{C00000} 149.83 }        & {\color[HTML]{E97132} 149.80}     & 149.81    & 0.02        \\
\ce{O2}           &118.03          & {\color[HTML]{C00000} 118.03 }        & {\color[HTML]{E97132} 117.96}     & 117.99    & 0.00        \\
\ce{OF}           & 51.13          & {\color[HTML]{C00000}  51.12 }         & {\color[HTML]{E97132} 51.17}      & 51.10     & 0.03        \\
\ce{F2}           & 37.01          & {\color[HTML]{C00000}  36.99 }         & {\color[HTML]{E97132} 36.95}      & 36.93     & 0.00        \\
\ce{PH3}          &227.33          & {\color[HTML]{C00000} 227.31 }        & {\color[HTML]{E97132} 227.36}     &           &             \\
\ce{HS}           & 83.66          & {\color[HTML]{C00000}  83.65 }         & {\color[HTML]{E97132} 83.68}      & 83.67     & 0.05        \\
\ce{H2S}          &173.59          & {\color[HTML]{C00000} 173.59 }        & {\color[HTML]{E97132} 173.58}     & 173.58    & 0.05        \\
\ce{HCl}          &102.26          & {\color[HTML]{C00000} 102.26 }        & {\color[HTML]{E97132} 102.20}     & 102.21    &  0.00           \\

\ce{SO}           &123.80          & {\color[HTML]{C00000} 123.77 }        & {\color[HTML]{E97132} 123.70}     & 123.74    & 0.05        \\
\ce{ClO}          & 63.46          & {\color[HTML]{C00000}  63.47 }         & {\color[HTML]{E97132} 63.37}      & 63.42     & 0.01        \\
\ce{ClF}          & 60.38          & {\color[HTML]{C00000}  60.37 }         & {\color[HTML]{E97132} 60.28}      & 60.35     & 0.01        \\
\ce{P2}           &116.18          & {\color[HTML]{C00000} 116.23 }        & {\color[HTML]{E97132} 116.22}     &   116.05$^a$        &   0.09$^a$          \\

\ce{S2}           &101.92          & {\color[HTML]{C00000} 101.93 }        & {\color[HTML]{E97132} 101.77}     & 101.89    & 0.07$^b$        \\
\ce{Cl2}          & 57.21          & {\color[HTML]{C00000}  57.22 }         & {\color[HTML]{E97132} 57.07}      & 57.18     & 0.00 \\      \hline

Extras:\\
\hline
\ce{SiF4} & 566.15 &          &        {\color[HTML]{E97132} 565.92}      & 565.99     & 0.18$^{e}$ \\
\ce{SiH3F}   &  363.93 & 363.92         &        {\color[HTML]{E97132} 363.69}      & 363.75     & 0.22$^{e}$ \\
\hline\hline
\end{tabular}
}

(*) Averaging weighted by $1/\max{(\mathrm{uncertainty,0.01})}$, excluding FOOF and OCS.

(a) From Gurvich\cite{Gurvich1989}, page 399.

(b) Spectroscopic determination from Frederix et al.\cite{Frederix2009}: $D_0$(\ce{S2})=35636.9$\pm$2.5 cm$^{-1}$, or 101.89$\pm$0.01 kcal/mol

(c) 
Erroneous original\cite{jmlm273} value corrected in Ref.\cite{jmlm330}

(d) DBOC omitted due to poles (see text).

(e) Provisional ATcT values.

(f) Original value\cite{jmlm273} had erroneous CV correction (see text)

\end{table}

For the W4-08 subset, old W4 (and for smaller species, W4.3 or W4.4) TAE$_0$ values taken from the W4-17 database\cite{jmlm273} are compared in Table~\ref{tab:ATcT} 
with presently obtained W5prelim1 and W5prelim2 values, as well as with ATcT (Active Thermochemical Tables\cite{ATcTreview}) version 1.220 (the most recent version as of December 21, 2025).\cite{ATcT} For some experimentally well-established species where significant gaps existed between W4-17 and ATcT, the gap is now closed smoothly, for example for \ce{Cl2}, \ce{S2} 
(see the spectroscopic dissociation energy of Frederix et al.\cite{Frederix2009}),  \ce{P2} (see Gurvich\cite{Gurvich1989}) and \ce{CN} radical. 

A mild degree of error cancellation exists between various components of W4-17. Yet for some second-row molecules, differences between old W4-17 and present data reach or exceed 0.30 kcal/mol. While we have reason to believe the new data stand on a more solid theoretical foundation, this is hard to know for sure without re-iterating the (at this point gargantuan) ATcT\cite{ATcTpaper1,ATcTreview} thermochemical network.

A significant gap between old and new values is also seen for boron hydrides, ethane, and the like. As these species are quite well-behaved from an electronic structure point of view, this observation results almost entirely from correlation contributions to the DBOC, which were neglected in the W4-17 work but become quite significant for these species.

\subsection{A remark on some atomic heats of formation}

Atomic heats of formation in the gas phase are required whenever a computed TAE$_0$ is to be converted into a molecular heat of formation in the gas phase. The current version of ATcT contains slight revisions of $\Delta H^\circ_{f,0 K}[A(g)]$ for A=\{C,N\}, and a more drastic revision for boron, where $\Delta H^\circ_{f,0 K}[B(g)]$ went up\cite{Ruscic_Bross_2025_boron} from 133.82$\pm$1.20 to 135.129$\pm$0.146 kcal/mol. The latter was consistent with earlier (Ref.\cite{jmlm204} and references therein) predictions extracted from experimental \emph{molecular} heats of formation and computed total atomization energies.
For instance, $\Delta H^\circ_{f,0 K}[\mathrm{Al(g)}]=\Delta H^\circ_{f,0 K}[\mathrm{AlF_3(g)}]+\mathrm{TAE_0}[\mathrm{AlF_3}]-3\Delta H^\circ_{f,0 K}[\mathrm{F(g)}]$.
Our calculations for \ce{BF3}, \ce{B2H6}, etc. indicate that the new ATcT value for boron atom is reliable within the stated uncertainty. 

Very recently [D. E. Bross and B. Ruscic, to be published], data for silicon compounds have been added to ATcT 1.220. The ATcT $\Delta H^\circ_{f,0 K}[\mathrm{Si(g)}]$=107.63$\pm$0.13 kcal/mol is considerably higher than the CODATA and Gurvich values of 106.5$\pm$1.9 and 106.6$\pm$1.9 kcal/mol, respectively, albeit within their very wide error bars. This follows Martin and Taylor\cite{jmlm126} proposing a somewhat milder upward revision to 107.15$\pm$0.38 kcal/mol about 25 years ago based on combining a CCSD(T)/AV\{Q,5\}Z+2d1f\hspace{0.04cm}   TAE$_0$ corrected for relativity and core-valence correlation --- at the time a calculation that required a powerful supercomputer --- with a very accurate experimental (fluorine bomb calorimetry) $\Delta H^\circ_{f,0 K}[\ce{SiF4(g)}]$. A followup paper\cite{jmlm204}  at the W4 level, averaging over values extracted from \ce{SiF4} and \ce{Si2H6}, slightly revised this to 107.2$\pm$0.15 kcal/mol. Our computed TAE$_0$ values for SiH, \ce{SiH4}, \ce{Si2H6}, SiO, SiF, and \ce{SiF4} are all within their ATcT error bars, and in fact suggest that the latter may be overly conservative.

For Al(g), the Gurvich\cite{GurvichVol3} $\Delta H^\circ_{f,0 K}[\mathrm{AlF_3(g)}]$=-288.13$\pm$0.74 kcal/mol and the Konings and Booij\cite{KoningsBooij1992_AlCl3} $\Delta H^\circ_{f,0 K}[\mathrm{AlCl_3(g)}]$=-139.57$\pm$0.43 kcal/mol, combined with ATcT heats of formation of F and Cl, and our present W5prelim2 TAE$_0$ values, lead to $\Delta H^\circ_{f,0 K}[\mathrm{Al(g)}]$=80.20$\pm$0.74 and 80.56$\pm$0.43 kcal/mol, respectively. These agree to overlapping uncertainties with each other and with the recommendation of Karton and Martin\cite{jmlm207}, 80.2$\pm$0.4 kcal/mol, which represented a ca. 2 kcal/mol upward revision from the CODATA value\cite{CODATA_Web} of 78.30$\pm$0.96 kcal/mol.

For phosphorus, we have $\Delta H^\circ_{f,0 K}[\mathrm{P(g)}]=(\Delta H^\circ_{f,0 K}[\mathrm{P_4(g)}]+\mathrm{TAE_0}[\mathrm{P_4}])/4$, which, with the CODATA $\Delta H^\circ_{f,0 K}[\mathrm{P_4(g)}]=15.03\pm0.07$ kcal/mol, leads to 
$\Delta H^\circ_{f,0 K}[\mathrm{P(g)}]$=75.45 kcal/mol with the W4-17 data, 75.51 kcal/mol for W5prelim2, and 75.52 kcal/mol for W5prelim1. These values corroborate an earlier suggestion\cite{jmlm207} that the CODATA value of 75.45$\pm$0.24 kcal/mol is substantially correct.

\section{Conclusion}

We have considered here, in the context of accurate thermochemistry, the role of subvalence correlation in two respects: (a) improved reference geometry; (b) post-CCSD(T) subvalence contributions to the atomization energy.

We find their effects to be comparatively mild for most first-row systems (more significant where there is strong static correlation). For second-row systems, however --- especially those with adjacent second-row atoms like \ce{S3}, \ce{S4}, \ce{P2}, and \ce{P4} --- contributions can get quite nontrivial, exceeding 0.5 kcal/mol for \ce{S4} (which is a `double whammy' in also having strong static correlation). As such, while the older W4 and W4.3/W4.4 are still acceptably reliable for first-row compounds. 

We also propose here two first attempts at a W5 protocol, and present revised TAE$_0$ values for the W4-08 subset, plus \ce{SiF4} and \ce{SiH3F}. Our predicted TAE$_0$ values (TAE at 0 K) agree well with the ATcT (active thermochemical tables) values, including for the very recent expansion of the ATcT network to boron, silicon, and sulfur compounds.

As subsidiary points, ROCCSD(T) optimized geometries are definitely preferred over UCCSD(T) for radicals (especially those with significant spin contamination), and the `averaged extrapolations' approach of Thorpe et al. is fortuitously in good agreement with energy optimization.

\begin{acknowledgments}
JMLM would like to acknowledge inspiring discussions at the Quantum Theory Project, U. of Florida, and at the weekly ATcT (Active Thermochemical Tables) Zoom meetings, with Drs. James H. Thorpe (QTP, now Argonne), Peter W. Franke (QTP), Megan R. Bentley  (QTP), David H. Bross (Argonne), and especially the hosts Dr. Branko Ruscic (Argonne) and Prof. John F. Stanton (1961-2025), of blessed memory. GHJ was funded by the National Science Foundation (grant CHE-2430408, ``Advances in Coupled Cluster Theory''; deceased PI: John F. Stanton; current PI: Alberto Perez). Research at Weizmann was supported in part by a grant to the Martin group from the Minerva Foundation, Munich, Germany, and by a generous allocation of computer time on the Faculty of Chemistry's HPC facility CHEMFARM. The latter acknowledges support from the Ben May Center for Chemical Theory and Computation, Weizmann Institute of Science.
\end{acknowledgments}

\section*{CRediT authorship contribution statement}

AB: simulations (lead), data curation, analysis (supporting), draft (supporting), ToC graphic (design and drawing);
GHJ: coding in CFOUR, methodology (supporting), analysis (supporting), draft (supporting);
KEW: simulations (relativistic corrections and DBOC), data curation, draft (supporting);
MS: funding acquisition (supporting), methodology (supporting), analysis (supporting), draft (supporting), funding acquisition (supporting);
JMLM: conceptualization, methodology (lead), data curation, analysis (lead), draft (lead), funding acquisition (lead). 
All authors contributed to writing – review and editing.

\subsection*{Supporting information}

Spreadsheet in Microsoft Excel format with all relevant data; reoptimized geometries in .xyz format for nearly all W4-17 species at the frozen-core CCSD(T)/cc-p(Q+d)Z and subvalence-correlated CCSD(T)/cc-pwCVQZ levels.

\bibliographystyle{achemso}
\bibliography{CVpostT,theRest}

\clearpage
\begingroup
\renewcommand\thefigure{} 
\renewcommand\figurename{} 
\setlength{\fboxsep}{4pt}    
\setlength{\fboxrule}{0.6pt} 

\begin{figure}[p]
  \centering
  \fbox{\includegraphics[width=0.9\linewidth]{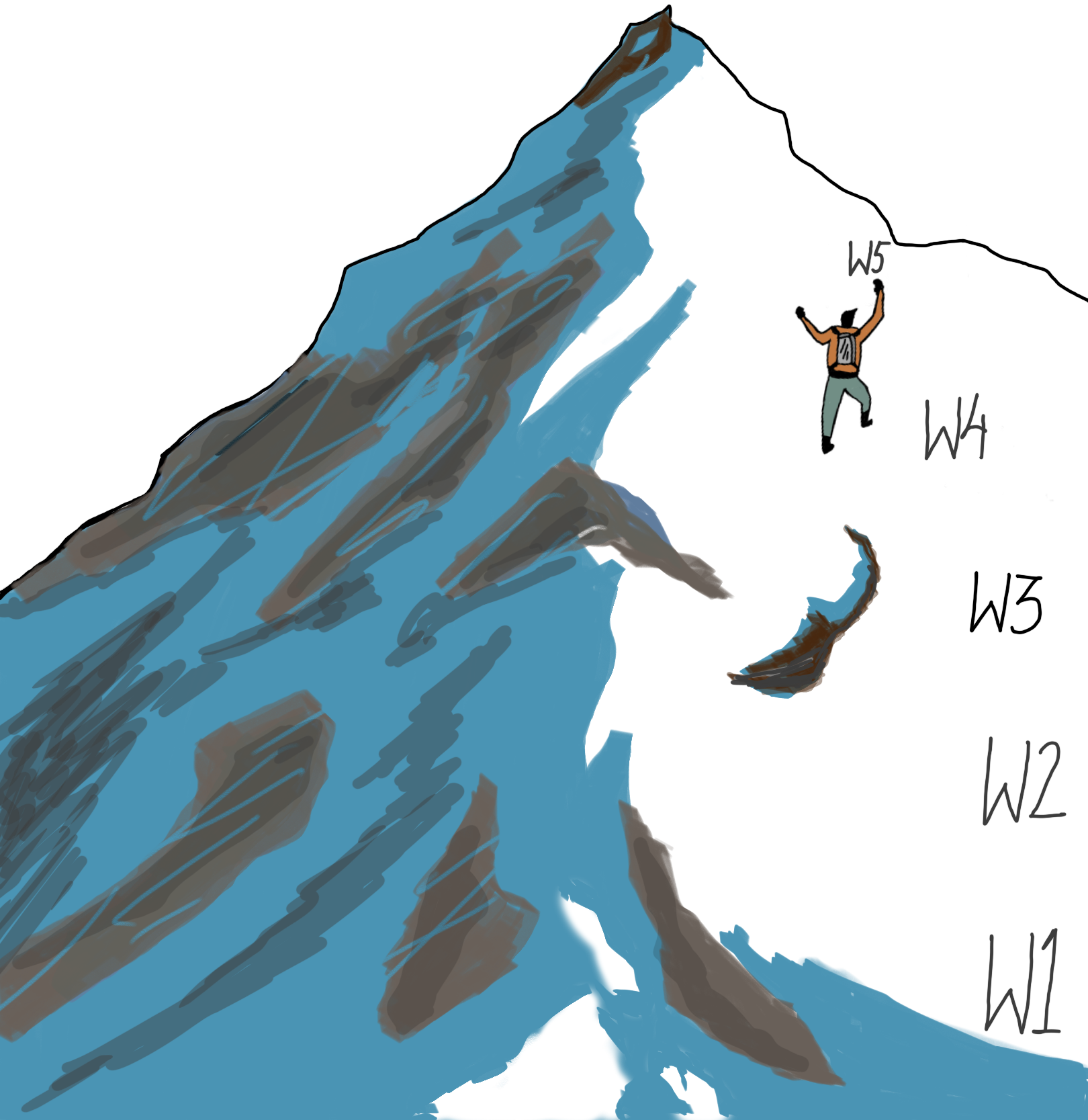}}
  \caption{Table of Contents Graphic}
\end{figure}
\endgroup

\end{document}